\documentclass[a4paper,10pt]{article}
\usepackage{graphicx}
\usepackage{comment}
\usepackage{amssymb}
\usepackage{amsmath}
\usepackage{nicefrac}
\usepackage{graphicx}
\usepackage{dcolumn}
\usepackage{bm}
\usepackage{comment}
\usepackage{multirow}
\usepackage{cite}
\usepackage{xcolor}
\usepackage{url}
\usepackage{mathrsfs}
\usepackage{supertabular}
\usepackage[normalem]{ulem}
\usepackage{lscape}
\usepackage{soul}
\usepackage{natbib}
\usepackage{subcaption} 
\usepackage{bold-extra}
\usepackage{hyperref}

\begin{document}
            
\newcommand{\bin}[2]{\left(\begin{array}{c}\!#1\!\\\!#2\!\end{array}\right)}
\newcommand{\threej}[6]{\left(\begin{array}{ccc}#1 & #2 & #3 \\ #4 & #5 & #6 \end{array}\right)}
\newcommand{\sixj}[6]{\left\{\begin{array}{ccc}#1 & #2 & #3 \\ #4 & #5 & #6 \end{array}\right\}}
\newcommand{\regge}[9]{\left[\begin{array}{ccc}#1 & #2 & #3 \\ #4 & #5 & #6 \\ #7 & #8 & #9 \end{array}\right]}
\newcommand{\La}[6]{\left[\begin{array}{ccc}#1 & #2 & #3 \\ #4 & #5 & #6 \end{array}\right]}
\newcommand{\hj}{\hat{J}}
\newcommand{\hux}{\hat{J}_{1x}}
\newcommand{\hdx}{\hat{J}_{2x}}
\newcommand{\huy}{\hat{J}_{1y}}
\newcommand{\hdy}{\hat{J}_{2y}}
\newcommand{\huz}{\hat{J}_{1z}}
\newcommand{\hdz}{\hat{J}_{2z}}
\newcommand{\hup}{\hat{J}_1^+}
\newcommand{\hum}{\hat{J}_1^-}
\newcommand{\hdp}{\hat{J}_2^+}
\newcommand{\hdm}{\hat{J}_2^-}

\huge

\begin{center}
Corrections to radiative rates between atomic configurations
\end{center}

\vspace{0.5cm}

\large

\begin{center}
Jean-Christophe Pain$^{a,b,}$\footnote{jean-christophe.pain@cea.fr} and Djamel Benredjem$^{c}$
\end{center}

\normalsize

\begin{center}
\it $^a$CEA, DAM, DIF, F-91297 Arpajon, France\\
\it $^b$Universit\'e Paris-Saclay, CEA, Laboratoire Mati\`ere en Conditions Extr\^emes,\\
\it 91680 Bruy\`eres-le-Ch\^atel, France\\
\it $^c$Laboratoire Aim\'e Cotton, Universit\'e Paris-Saclay, Orsay, France\\
\end{center}

\vspace{0.5cm}

\begin{abstract}
The computation of radiative opacity or emissivity of hot dense matter is a challenging task. It requires accounting for an immense number of energy levels and lines across various excitation and ionization states. Whether in local thermodynamic equilibrium (LTE) or non-LTE plasmas, statistical methods provide significant assistance. Many computational codes are based on the Detailed Configuration Accounting approximation, which involves averaged rates between configurations. In that approach, only the mean energies of the configurations are considered, and the effects of the energy distributions of the levels within the initial and final configurations are typically neglected. A long time ago, Klapisch proposed a method to correct the rates. The corresponding formalism includes the energy shift and variance of the Unresolved Transition Array, as well as the average energies of the configurations. We extend this formalism and investigate its impact on opacity calculations in two specific cases: first, the iron experiment conducted at Sandia National Laboratories under conditions similar to those at the base of the Sun's convective zone, dominated by L-shell 2p-$n$d transitions, and second, laser experiments--still for iron--at much lower temperature. The latter measurements shed light on our understanding of the envelopes of $\beta$-Cephei-type stars, where the relevant transitions are intra-M-shell $\Delta n=0$ (3-3) transitions, specifically 3s-3p and 3p-3d, in the XUV range. The issue of ensuring the validity of Kirchhoff's law when plasmas approach LTE is also addressed, and a prescription is proposed, applying both to the standard configuration-to-configuration case and to the aforementioned corrections, which account for the energy distribution of the levels within a configuration.
\end{abstract}

\section{Introduction}\label{sec0}

The field of atomic physics in plasmas emerged alongside astrophysics, as the latter developed distinctly from classical astronomy. Early applications focused on modeling optically thick plasmas, such as those found in stellar interiors and atmospheres, as well as optically thin, low-density plasmas emitting X-rays, like solar and stellar coronas. These challenges spurred significant advancements in both atomic structure theory and the theory of electron-ion and proton-ion collisions. As laboratory plasma devices were later developed, physicists began leveraging theoretical frameworks originally created for astrophysical contexts. Separately, extensive atomic data—such as energy levels, oscillator strengths, and cross-sections—had long been obtained through conventional atomic physics experiments, independent of plasma-generating setups. Continued progress in this area now relies on sophisticated tools like electron beam ion traps and storage rings. In the domain of plasma physics proper, cutting-edge experiments today use lasers or magnetic confinement devices such as tokamaks, Z-pinches, and stellarators. A new wave of inquiry is also driven by experiments using X-ray free-electron lasers, which require complex theoretical modeling for proper interpretation.

In high-temperature plasmas, atoms and ions undergo interactions with free electrons and photons, potentially populating an enormous number of quantum states—often in the millions. This complexity makes direct tracking of all transitions impractical. To manage this, collective or global modeling techniques are employed. For instance, a line spectrum can be treated statistically by viewing it as a distribution of line wavenumbers weighted by their intensities. One can then compute its mean energy and width—the first two statistical moments—using radial integrals, enabling the spectrum to be approximated as a Gaussian distribution, thereby avoiding the need to diagonalize large energy matrices.

Similarly, the challenge of determining level populations can be addressed by grouping numerous quantum levels into manageable ensembles—such as configurations or even superconfigurations. This allows for the construction of a reduced system of rate equations to compute their populations. Thus, by integrating methods from classical statistics with those of atomic physics, it is possible to derive meaningful global properties—such as average energies or line intensities—of vast ensembles of atomic states, without explicitly solving for every individual basis state.

Thus, absorption and emission of plasmas composed of mid- to high-$Z$ elements present broadband line emission features called unresolved transition arrays \cite{Mandelbaum1987,Mandelbaum1991,Mandelbaum1994}. Each of these groups of lines (transition arrays) corresponds to the whole transition set between two specific configurations $C$ and $C'$. To analyze an Unresolved Transition Array (UTA) using a statistical method, Bauche, Bauche-Arnoult and Klapisch evaluated the strength-weighted distribution of its spectral line wavenumbers \cite{Bauche1979}. The first statistical moment, representing the strength-weighted mean of these wavenumbers, is expressed as a linear combination of various radial integrals within the Hamiltonian framework. These include the direct and exchange Slater integrals ($F^k$ and $G^k$) associated with electron-electron electrostatic repulsion, as well as the spin-orbit interaction integrals. The second moment, which is related to the full width at half maximum (FWHM) of the distribution, involves linear combinations of both squared terms and cross-products of the Slater and spin-orbit integrals.

The definition of configuration-averaged rates has been the subject of a number of investigations, especially in the development of collisional-radiative models (see for instance \cite{Peyrusse1999,Peyrusse2000,Hansen2007,Abdallah2008,Hansen2011}. Recently, we investigated the ionization and excitation processes induced by electron impact between two configurations or superconfigurations. In that work, rate coefficients are calculated for transition arrays or super-transition arrays rather than level-to-level transitions. Special attention is given to a series of oxygen-like ions relevant to inertial confinement fusion, specifically silicon, germanium, argon, and krypton \cite{Benredjem2025}. In the present work, we concentrate on the spontaneous-emission rate.

In the next section, we present a method of calculating the configuration-averaged rates as a first approximation. In section \ref{sec2}, we discuss corrections to the rates. In section \ref{sec3}, the impact of these corrections in the case of an iron plasma at $\rho$ = 0.17 g/cm$^3$ and T = 182 eV is investigated. Such conditions correspond to the boundary of the convective zone of the Sun, and involve mostly 2p-$n$d, $n\geq 3$ transitions. The magnitude of the corrections is studied in Sec. \ref{sec4} for an iron plasma at $\rho$ = 0.01 g/cm$^3$ and T = 22 eV. With respect to temperature, these conditions are similar to those occurring in the envelopes of $\beta$-Cephei type stars (as concerns the mean ionization), and are typical of laser photo-absorption measurements \cite{Winhart1996,Turck2016}. In the latter case, the important transitions are mostly $\Delta n=0$ (3-3) ones. The issue of preserving the detailed balance with configuration-averaged rates is addressed in Sec. \ref{sec5}, and a procedure ensuring the validity of Kirchoff's law is proposed.

\section{Configuration-averaged rate coefficients}\label{sec1}

\subsection{General formulation}\label{subsec11}

If $J$ is the total atomic angular momentum and $\alpha$ the ensemble of additional quantum numbers required to label a level unambiguously, a rate coefficient connecting two levels $\alpha J\in C$ and $\alpha'J'\in C'$ reads
\begin{equation*}
    T_{\alpha J,\alpha'J'}=\sum_k\mathscr{T}_k(\alpha J,\alpha'J')\,\Theta_k(E_{\alpha J,\alpha'J'}),
\end{equation*}
where $\mathscr{T}_k$ is the square of a purely angular, energy-independent matrix element of a tensor operator of rank $k$, and the energy dependence is relegated to the radial factor $\Theta$. One has $E_{\alpha J,\alpha' J'}=E_{\alpha J}-E_{\alpha' J'}$, where $E_{\alpha J}$ is the energy of level $\alpha J$. Actually, the configuration-averaged rate is obtained as a sum over levels of the final configuration $C'$, and averaged over levels of the initial configuration $C$:
\begin{equation*}
    T_{CC'}=\sum_{\alpha'J'\in C'}\langle T_{\alpha J,\alpha'J'}\rangle_{\alpha J\in C}
\end{equation*}
with
\begin{equation*}
    \langle T_{\alpha J,\alpha'J'}\rangle_{\alpha J\in C}=\frac{\sum_{\alpha J\in C}(2J+1)\,e^{-\beta E_{\alpha J}}\,T_{\alpha J,\alpha'J'}}{\sum_{\alpha J\in C}(2J+1)\,e^{-\beta E_{\alpha J}}}, 
\end{equation*}
where $\beta=1/(k_BT)$, $k_B$ being the Boltzmann constant. In the high-temperature limit ($\beta\rightarrow 0$), all the exponentials are equal to one, and one gets
\begin{equation*}
    T_{CC'}=\frac{1}{g_C}\sum_{(\alpha J,\alpha'J') \in C\otimes C'}(2J+1)\,T_{\alpha J,\alpha'J'},
\end{equation*}
where $(\alpha J,\alpha'J') \in C\otimes C'$ means that $\alpha J\in C$ and $\alpha' J'\in C'$ are the initial and final levels of an electric-dipole line (according to the corresponding selection rules $J'=J\pm 1$ or $J'=J\ne 0$), and
\begin{equation*}
    g_C=\sum_{\alpha J\in C}(2J+1)
\end{equation*}
is the degeneracy of configuration $C$. Defining and calculating rates is therefore not an innocent matter, and is often subject to approximation(s) \cite{Poirier2007,Poirier2008}. In particular $E_{\alpha J,\alpha' J'}$ is often replaced by differences between the energies of configurations $C$ and $C'$.

\subsection{Level-to-level Einstein equations and configuration-averaged radiative rates}\label{subsec12}

To simplify the notations, we replace $\alpha J$ by $u$ and $\alpha' J'$ by $d$ in the following. The spontaneous-emission coefficient $A_{ud}$ between levels $u$ and $d$ is related to the stimulated emission coefficient $B_{ud}$ by
\begin{equation}\label{un}
    \frac{A_{ud}}{B_{ud}}=\frac{2h\nu_{du}^3}{c^2},
\end{equation}
where $h\nu_{du}=E_u-E_d$, and the absorption coefficient $B_{du}$ is related to $B_{ud}$ by
\begin{equation}\label{deux}
    \frac{B_{du}}{B_{ud}}=\frac{g_u}{g_d},
\end{equation}
where $g_u$ and $g_d$ are the degeneracies of levels $u$ and $d$ respectively. We also have
\begin{equation}\label{tro}
    A_{ud}=\frac{16\pi^3\left(E_u-E_d\right)^3}{3h^4\epsilon_0c^3g_u}S_{ud}
\end{equation}
where $S_{ud}$ is the line strength. Thus, according to Eq. (\ref{un}):
\begin{equation*}
    B_{ud}=\frac{8\pi^3}{3h^2\epsilon_0c\,g_u}S_{ud}
\end{equation*}
and according to Eq. (\ref{deux}):
\begin{equation*}
    B_{du}=\frac{8\pi^3}{3h^2\epsilon_0c\,g_d}S_{ud}
\end{equation*}
since $S_{ud}=S_{du}$.

Let us set $A_{ud}=C_0\left(E_u-E_d\right)^3S_{ud}/g_u$ and $B_{ud}=C_0h^2c^2S_{ud}/(2g_u)$, with $C_0=16\pi^3/(3h^4\epsilon_0c^3)$. One has, for the spontaneous-emission rate between configurations $C$ and $C'$:
\begin{equation*}
    A_{CC'}=\frac{\sum_{(u,d)\in C\otimes C'}A_{ud}\,g_u\,e^{-\beta E_u}}{\sum_{u\in C}g_u\,e^{-\beta E_u}},
\end{equation*}
which, according to Eq. (\ref{tro}), results in
\begin{equation*}
    A_{CC'}=C_0\frac{\sum_{(u,d)\in C\otimes C'}S_{ud}\left(E_u-E_d\right)^3\,e^{-\beta E_u}}{\sum_{u\in C}\,g_u\,e^{-\beta E_u}},
\end{equation*}
that can be approximated by
\begin{equation*}
    A_{CC'}\approx C_0\frac{\sum_{(u,d)\in C\otimes C'}S_{ud}\,e^{-\beta E_u}}{\sum_{u\in C}g_u\,e^{-\beta E_u}}\frac{\sum_{(u,d)\in C\otimes C'}S_{ud}\left(E_u-E_d\right)^3}{\sum_{(u,d)\in C\otimes C'}S_{ud}}
\end{equation*}
and finally:
\begin{equation}\label{avecmu3}
    A_{CC'}\approx C_0\frac{S_{C'C}}{g_{C'}}e^{-\delta E_{C'}}\mu_3,
\end{equation}
where $\mu_3$ is the third-order moment of the lines and $\delta E_{C'}$ the difference between the average level energies weighted by the line strength and the average level energies weighted by the degeneracies, for the configuration $C'$. Setting $E_u-E_d=E_{ud}=E$ and noting that
\begin{equation*}
    E^3=\langle E\rangle^3+3\left(E-\langle E\rangle\right)\langle E\rangle^2+3\left(E-\langle E\rangle\right)^2\langle E\rangle+\left(E-\langle E\rangle\right)^3,
\end{equation*}
a fair approximation to $\mu_3$ is given by
\begin{equation*}
    \mu_3\approx\mu_1^3\left(1+3\frac{\sigma_{CC'}^2}{\mu_1^2}\right),
\end{equation*}
where $\mu_1=E_{CC'}+\delta E_{CC'}$ (see Eq. (\ref{mu1rap})), $E_{CC'}$ being the transition energy $E_C-E_{C'}$ ($E_C$ and $E_{C'}$ are the average energies of configurations $C$ and $C'$ respectively). In a similar way, we get
\begin{equation*}
    B_{CC'}=\frac{C_0h^2c^2}{2}\frac{\sum_{d\rightarrow u}S_{ud}\,e^{-\beta E_d}}{\sum_dg_de^{-\beta E_d}}
\end{equation*}
\begin{equation*}
    B_{CC'}=\frac{C_0h^2c^2}{2}\frac{\sum_{(u,d)\in C\otimes C'}S_{ud}\,e^{-\beta E_d}}{\sum_{d\in C}g_de^{-\beta E_d}}
\end{equation*}
and thus 
\begin{equation*}
    B_{CC'}\approx\frac{C_0h^2c^2}{2}\frac{S_{C'C}}{g_{C}}e^{-\delta E_{C}},
\end{equation*}
where $\delta E_{C}$ is the difference between the average level energies weighted by the line strength and the average level energies weighted by the degeneracies, for the configuration $C$. Similarly, one has
\begin{equation*}
    B_{C'C}\approx\frac{C_0h^2c^2}{2}\frac{S_{C'C}}{g_{C'}}e^{-\delta E_{C'}}.
\end{equation*}

\section{Corrections to the rate coefficients}\label{sec2}

\subsection{Series expansions}\label{subsec21}

Expanding a $\Theta$ function as a Taylor series, one gets
\begin{equation*}
    \Theta_k(E_{ud})=\Theta_k(E_{CC'})+\sum_{n=1}^{\infty}\frac{(E_{ud}-E_{CC'})^n}{n!}\,\displaystyle\frac{\partial^n\Theta_k}{\partial E^n}\Big|_{E_{CC'}}.
\end{equation*}
In particular
\begin{align*}
    \Theta_k(E_{ud})=&\Theta_k(E_{CC'})+(E_{ud}-E_{CC'})\,\displaystyle\frac{\partial\Theta_k}{\partial E}\Big|_{E_{CC'}}+\frac{1}{2}(E_{ud}-E_{CC'})^2\,\displaystyle\frac{\partial^2\Theta_k}{\partial E^2}\Big|_{E_{CC'}}\nonumber\\
    &+\frac{1}{6}(E_{ud}-E_{CC'})^3\,\displaystyle\frac{\partial^3\Theta_k}{\partial E^3}\Big|_{E_{CC'}}+O\left((E_{ud}-E_{\mathrm{CC'}})^3\right).
\end{align*}
The rate between configurations $C$ and $C'$ is
\begin{equation}\label{expan}
    T_{CC'}=\frac{1}{g_C}\sum_k\left\{\mathcal{S}(\mathscr{T}_k)\,\Theta_k(E_{CC'})\left[1+\mu_1^{(k)}\frac{1}{\Theta_k}\frac{\partial\Theta_k}{\partial E}\Big|_{E_{CC'}}+\frac{\mu_2^{(k)}}{2}\frac{1}{\Theta_k}\frac{\partial^2\Theta_k}{\partial E^2}\Big|_{E_{CC'}}+\frac{\mu_3^{(k)}}{6}\frac{1}{\Theta_k}\frac{\partial^3\Theta_k}{\partial E^3}\Big|_{E_{CC'}}+\cdots\right]\right\},
\end{equation}
where $\mathcal{S}$ is a function of $\mathscr{T}_k$ obtained by usual sum rules, and $\mu_n^{(k)}$ are the generalized UTA moments (restoring the notations $u\rightarrow \alpha J$ and $d\rightarrow \alpha' J'$):
\begin{equation}\label{mun}
    \mu_n^{(k)}=\frac{\sum_{\alpha J,\alpha'J'}\left(E_{\alpha J}-E_{\alpha' J'}\right)^n\,\mathscr{T}_k(\alpha J,\alpha'J')}{\sum_{\alpha J,\alpha'J'}\mathscr{T}_k(\alpha J,\alpha'J')}
\end{equation}
involving potentially other operators than the electric dipole. The above formalism can be applied to radiative rates, such as the Einstein spontaneous-emission coefficient $A$, which is the focus of this work. The method can also be extended to collisional-excitation. In this case, the collisional-excitation rate coefficient can be expressed as a sum of multipole one-electron operators acting only on the atomic bound electrons \cite{Barshalom1988b}. Such operators are multiplied by radial integrals $Q_k$ that include the sum over all partial waves. The collision strength $\Omega_{CC'}$ between configurations $C$ and $C'$ can be expressed as
\begin{equation*}
    \Omega_{CC'}=\frac{1}{g_C}\sum_k\left\{\mathscr{S}_k\,\Theta_k(E_{CC'})\left[1+\mu_1^{(k)}\frac{1}{Q_k}\frac{\partial Q_k}{\partial E}\Big|_{E_{CC'}}+\frac{\mu_2^{(k)}}{2}\frac{1}{Q_k}\frac{\partial^2Q_k}{\partial E^2}\Big|_{E_{CC'}}+\frac{\mu_3^{(k)}}{6}\frac{1}{Q_k}\frac{\partial^3Q_k}{\partial E^3}\Big|_{E_{CC'}}+\cdots\right]\right\},
\end{equation*}
where $\mathscr{S}_k$ represents an angular sum rule for collision strength and $\theta_k$ is a radial factor. The formulas for the moments of the multipoles can be derived with the same methods that were used for the electric-dipole UTA, in the case of radiative rates \cite{Bauche1988}, and it turns out that the $Q_k$ vary almost linearly with the transition energy over a wide range (except for $k = 1$, for which the second-order correction cannot be neglected).\\

In the following we focus on corrections to the spontaneous-emission coefficient $A$.

\subsection{Corrections up to second order for radiative rates: UTA shift and variance}\label{subsec22}

Assuming relevance for population kinetics, Peyrusse limited the expansion in Eq. (\ref{expan}) to the first two terms \cite{Peyrusse1999}. This assumption breaks down when $C$ and $C'$ are close to each other in energy. In that case, local thermodynamic equilibrium (LTE) dominates and precision for a given rate becomes less critical.

Focusing on the Einstein spontaneous-emission coefficient $A$, we have, for a one-electron jump from subshell $\alpha$ to subshell $\beta$:
\begin{equation*}
    A_{CC'}=a(E_C-E_{C'})^3\frac{n_{\alpha}(g_{\beta}-n_{\beta})}{g_{\alpha}g_{\beta}}2\left[\ell_{\alpha},\ell_{\beta}\right]\threej{\ell_{\beta}}{1}{\ell_{\alpha}}{0}{0}{0}^2P_{\alpha\beta}^2,
\end{equation*}
where $n_i$ and $g_i$ represent respectively the electron population and degeneracy of subshell $i$. We use the notation $[\ell]=2\ell+1$. $A_{CC'}$ is in s$^{-1}$, energies are in Rydberg, $a=2.677\times 10^9$, and
\begin{equation*}
    P_{\alpha\beta}=\int_0^{\infty}y_{n_{\alpha}\ell_{\alpha}}(r)\,y_{n_{\beta}\ell_{\beta}}(r)\,\mathrm{d}r,
\end{equation*}
$y_i$ being the radial part of the wavefunction of subshell $i$ multiplied by $r$. This formula works well as long as the energy spread of $\alpha J$ levels in configurations $C$ and $C'$ remains small in comparison with the energy difference $E_{CC'}$. In the opposite case, this is not true anymore. Applying the above corrections to electric dipole radiative transition yields:

\begin{equation}\label{ACC'-corr}
    A_{CC'}^{\mathrm{corr}}=A_{CC'}\left(1+3\frac{\delta E_{CC'}}{E_{CC'}}+3\frac{\sigma_{CC'}^2}{E_{CC'}^2}\right),
\end{equation}
where $\delta E_{CC'}$ and $\sigma_{CC'}^2$ are the UTA shift and the UTA variance, respectively. As pointed out by Klapisch \cite{Klapisch1993}, these corrections are important for $\Delta n=0$ transitions. For instance, for ionized rare earths with a 4d$^r$ ground configuration, the ratio $\delta E_{CC'}/(E_{CC'})$ can be more than 10 \% \cite{Mandelbaum1987}. The second-order correction is expected to be important for high-$n$ overlapping configurations. These corrections also influence the population of the first excited configuration $4d^{n-1}4f$ and therefore the ionization balance.

The moments (see Eq. (\ref{mun})) are now
\begin{equation}\label{mun2}
    \mu_n=\frac{\sum_{\alpha J,\alpha'J'}\left(E_{\alpha J}-E_{\alpha' J'}\right)^nA_{\alpha J,\alpha'J'}}{\sum_{\alpha J,\alpha'J'}A_{\alpha J,\alpha'J'}},
\end{equation}
and one has in particular \cite{Bauche1988,Bauche1994}:
\begin{equation}\label{mu1rap}
    \mu_1=E_C-E_{C'}+\delta E_{CC'}=E_{CC'}+\delta E_{CC'}.
\end{equation}
For transition arrays of the type $n\ell^{N+1}-n\ell^Nn'\ell'$, one has:
\begin{equation*}
    \delta E_{CC'}=N\frac{(2\ell+1)(2\ell'+1)}{(4\ell+1)}\left(\sum_{k\ne 0}f_k\,F^{(k)}(n\ell,n\ell')+\sum_kg_k\,G^{(k)}(n\ell,n\ell')\right),
\end{equation*}
where $f_k$ ad $g_k$ involve usual Wigner $3j$ and Racah $6j$ coefficients:
\begin{equation*}
    f_k=\threej{\ell}{k}{\ell}{0}{0}{0}\threej{\ell'}{k}{\ell'}{0}{0}{0}\sixj{\ell}{k}{\ell}{\ell'}{1}{\ell'}
\end{equation*}
and
\begin{equation*}
    g_k=\threej{\ell}{k}{\ell'}{0}{0}{0}^2\left(\frac{2}{3}\delta_{k,1}-\frac{1}{(2\ell+1)(2\ell'+1)}\right).
\end{equation*}
In the case of configurations $C=n\ell^{N+1}$ and $C'=n\ell^Nn'\ell'$ \cite{Klapisch1993}, the expression of the variance contains only squares of products of internal Slater integrals $F^k(\ell,\ell)$ (denoted $F_C^k$ and $F_{C'}^k$ in configurations $C$ and $C'$), and of external Slater integrals $F^k(\ell,\ell')$ and $G^k(\ell,\ell')$ (denoted $F^k$ and $G^k$ respectively), and of squares of cross products of spin-orbit integrals $\zeta_{\ell C}$, $\zeta_{\ell C'}$ and $\zeta_{\ell'}$. It can be shown that no cross products of Slater and spin-orbit integrals can occur. The largest contributions involve the internal Slater integrals. The dependencies $f_i(N)$ of the corresponding types of products $P_i$ with respect to $N$ are as follows
\begin{equation*}
\begin{array}{ll}
F_{C'}^kF_{C'}^{k'}: & f_1(N)=N(N-1)(4\ell-N+1)(4\ell-N+2)\\
F_C^kF_C^{k'}: & f_2(N)=N(N+1)(4\ell-N)(4\ell-N+1)\\
F_{C'}^kF_C^{k'}: & f_3(N)=-2N(N-1)(4\ell-N)(4\ell-N+1)\\
F_{C'}^kF^{k'} \text{ or } F_{C'}^kG^{k'}: & f_4(N)=N(N-1)(4\ell-N+1)\\
F_C^kF^{k'} \text{ or } F_{C}^kG^{k'} & f_5(N)=N(4\ell-N)(4\ell-N+1).
\end{array}
\end{equation*}

The coefficient in $\sigma^2$ of each factor $P_i$ is written as the product of its dependency $f_j(N)$ with respect to $N$ by a quantity $Q_i$ independent of $N$. For example, $\sigma_{CC'}^2$ contains, as concerns the internal parameters only, the sum
\begin{align*}
    &\sum_{k,k'}\left(2\frac{\delta_{k,k'}}{2k+1}-\frac{1}{(2\ell+1)(4\ell+1)}-\sixj{\ell}{\ell}{k}{\ell}{\ell}{k'}\right)\nonumber\\
    &\;\;\;\;\;\;\;\;\times\frac{(2\ell+1)^3}{8\ell(4\ell-1)(4\ell+1)}\threej{\ell}{k}{\ell}{0}{0}{0}^2\threej{\ell}{k'}{\ell}{0}{0}{0}^2\nonumber\\
    &\;\;\;\;\;\;\;\;\times\left[f_1(N)\,F_{C'}^k\,F_{C'}^{k'}+f_2(N)\,F_{C}^kF_{C}^{k'}+f_3(N)\,F_{C'}^kF_{C}^{k'}\right],
\end{align*}
where $k$ and $k'$ are even and run from 0 to $2\ell$. It can be noted that, in the simple case where $F_C^k=F_{C'}^k$, the squared bracket in this sum (third line of the previous equation) becomes proportional to $N(4\ell-N+1)F_{C'}^kF_{C'}^{k'}$.

\subsection{Towards higher orders}\label{subsec23}
The next (third-order) term in the expansion (Eq. (\ref{ACC'-corr})) yields the correcting factor
\begin{equation*}
    \mathscr{F}=1+3\frac{\delta E_{CC'}}{E_{CC'}}+3\frac{\sigma_{CC'}^2}{E_{CC'}^2}+\frac{\mu_3-3\mu_1\mu_2+2\mu_1^3}{E_{CC'}^3}.
\end{equation*}
The problem is that $\mu_3$ is not easily known (although some parts of it were published \cite{Karazija1991,Karazija1991b,Karazija1995}). Therefore,
the ``direct'' calculation of the moments is complicated and the final result would be a very long formula. However, some algorithms have been proposed in order to evaluate these high-order moments. Karazija et al. \cite{Karazija1991,Karazija1991b,Karazija1995} expressed the spectral moments by averages of the products of operators and formulated a general group-diagrammatic method for the evaluation of their explicit expressions. Oreg et al. \cite{Oreg1990} considered the property that the moments reduce to configuration averages of $n$-boby symmetrical operators. For that purpose, the authors introduced the concept of an $n$-electron minimal configuration, relative to the actual ($N$-electron) configuration average. Their algorithm uses graphical technique (routine {\sc NJGRAF} \cite{Barshalom1988}) in order to derive the dependence of the averages on the orbital quantum numbers in terms of closed diagrams.

Regardless of how the configuration-averaged rates are calculated, it is important to note that in a non-LTE plasmas, detailed balance is no longer inherently satisfied. This issue must be addressed, as will be explained later.


\section{Case of an iron plasma at \texorpdfstring{$\rho=$0.17 g/cm$^3$}{rho=0.17 g/cm3} and \texorpdfstring{$T=$182 eV}{T=182 eV}}\label{sec3}

\subsection{Average atomic structure}\label{subsec31}

Nearly a hundred years ago, astronomers discovered that the way stellar material absorbs radiation determines the temperature structure within stars. For decades, no laboratory was able to replicate the extreme conditions of a stellar interior to directly measure these opacities, leaving stellar models with significant uncertainties. The issue became acute when refined analyses of the solar photosphere lowered the estimated carbon, nitrogen and oxygen abundances by 30–50 \%. Standard solar models built with these reduced abundances fail to match helioseismic data, which map the Sun's interior via acoustic oscillations. Raising the true average opacity of solar material by about 15 \% would counterbalance the lower element abundances and restore agreement. Iron is particularly important, contributing roughly one-quarter of the total opacity at the boundary between the radiative and convective zones. Bailey et al. have measured, on the Z-pinch machine facility of Sandia National Laboratories (SNL), iron's wavelength-resolved opacity at electron temperatures of 1.9 to 2.3 million K and electron densities of 0.7 to 4.0 $\times$ 10$^{22}$ cm$^{-3}$-conditions that closely mirror those at the solar radiative-convective interface, where the mismatch is largest \cite{Bailey2015}. Their results showed iron's opacity to be 30–400 \% higher across the spectrum than current theoretical models predict. Although iron is just one of several opacity-bearing elements, this excess accounts for roughly half of the additional mean opacity required to reconcile solar models with helioseismic observations.

Let us therefore consider an iron plasma at $\rho=0.17$ g/cm$^3$ and $T=$182 eV. Such conditions correspond to the SNL experiment and are typical of the base of the convective zone of the Sun \cite{Bailey2015,Buldgen2025}. The average energies of the subshells, computed with the average-atom model forming the first part (initialization) of our opacity code \cite{Pain2025}, are given in table \ref{tab1}. We can see that the last populated subshell at such a density is 8d.

\begin{table}[!ht]
\begin{subtable}[c]{0.5\textwidth}
\begin{tabular}{ccrr}\hline\hline
$n$ & $\ell$ & Energy & Population\\\hline\hline
1 & 0 & -0.7685$\times 10^{4}$ & 0.2000$\times 10^1$\\
2 & 0 & -0.1406$\times 10^{4}$ & 0.1637$\times 10^1$\\
2 & 1 & -0.1305$\times 10^{4}$ & 0.4328$\times 10^1$\\
3 & 0 & -0.4928$\times 10^{3}$ & 0.5790$\times 10^{-1}$\\
3 & 1 & -0.4613$\times 10^{3}$ & 0.1468\\
3 & 2 & -0.4212$\times 10^{3}$ & 0.1972\\
4 & 0 & -0.2228$\times 10^{3}$ & 0.1343$\times 10^{-1}$\\
4 & 1 & -0.2100$\times 10^{3}$ & 0.3758$\times 10^{-1}$\\
4 & 2 & -0.1943$\times 10^{3}$ & 0.5750$\times 10^{-1}$\\
4 & 3 & -0.1840$\times 10^{3}$ & 0.76085$\times 10^{-1}$\\
5 & 0 & -0.1086$\times 10^{3}$ & 0.7197$\times 10^{-2}$\\
5 & 1 & -0.1023$\times 10^{3}$ & 0.2086$\times 10^{-1}$\\
5 & 2 & -0.9460$\times 10^{2}$ & 0.3333$\times 10^{-1}$\\
5 & 3 & -0.8950$\times 10^{2}$ & 0.4537$\times 10^{-1}$\\
5 & 4 & -0.8695$\times 10^{2}$ & 0.57525$\times 10^{-1}$\\
6 & 0 & -0.5109$\times 10^{2}$ & 0.5252$\times 10^{-2}$\\\hline\hline
\end{tabular}
\end{subtable}
\begin{subtable}[c]{0.5\textwidth}
\begin{tabular}{ccrr}\hline\hline
$n$ & $\ell$ & Energy & Population\\\hline\hline
6 & 1 & -0.4758$\times 10^{2}$ & 0.15455$\times 10^{-1}$\\
6 & 2 & -0.4327$\times 10^{2}$ & 0.2516$\times 10^{-1}$\\
6 & 3 & -0.4029$\times 10^{2}$ & 0.3465$\times 10^{-1}$\\
6 & 4 & -0.3852$\times 10^{2}$ & 0.4412$\times 10^{-1}$\\
6 & 5 & -0.3719$\times 10^{2}$ & 0.5353$\times 10^{-1}$\\
7 & 0 & -0.1958$\times 10^{2}$ & 0.4419$\times 10^{-2}$\\
7 & 1 & -0.1752$\times 10^{2}$ & 0.1311$\times 10^{-1}$\\
7 & 2 & -0.1495$\times 10^{2}$ & 0.2154$\times 10^{-1}$\\
7 & 3 & -0.1301$\times 10^{2}$ & 0.2984$\times 10^{-1}$\\
7 & 4 & -0.1162$\times 10^{2}$ & 0.3807$\times 10^{-1}$\\
7 & 5 & -0.1034$\times 10^{2}$ & 0.4621$\times 10^{-1}$\\
7 & 6 & -0.8952$\times 10$ & 0.5419$\times 10^{-1}$\\
8 & 0 & -0.2786$\times 10$ & 0.4030$\times 10^{-2}$\\
8 & 1 & -0.1722$\times 10$ & 0.1202$\times 10^{-1}$\\
8 & 2 & -0.3481 & 0.1988$\times 10^{-1}$\\\hline\hline
\end{tabular}
\end{subtable}
\caption{Energies (in eV) and populations of subshells in an iron plasma at $\rho=$0.17 g/cm$^3$ and $T=$182 eV.}\label{tab1}
\end{table}

\subsection{2p-3d transitions}\label{subsec32}

Table \ref{tab2} provides the parameters of various 2p-3d transition arrays in an iron plasma at $\rho=$0.17 g/cm$^3$ and $T=$182 eV. The transition arrays are selected based on the code's initial output, specifically the first encountered 2p-3d transitions from the most probable initial configurations. The value in parentheses in the fifth column indicates the spin-orbit contribution to the variance, which is significant. The spin-orbit interaction causes the splitting of the $2p-3d$ into relativistic subarrays 2p$_{1/2}$-3d$_{3/2}$ and 2p$_{3/2}$-3d$_{5/2}$. Let us name the three corrections by
\begin{equation*}
    \delta f^{(1)}=3\frac{\delta E_{CC'}}{E_{CC'}}
\end{equation*}
for the first order,
\begin{equation*}
    \delta f^{(2)}=3\frac{\sigma_{CC'}^2}{E_{CC'}^2}
\end{equation*}
for the second order, and
\begin{equation*}
    \delta f^{(3)}=\frac{\mu_3-3\mu_1\mu_2+2\mu_1^3}{E_{CC'}^3}
\end{equation*}
for the third order. As can be seen in table \ref{tab2}, the second-order correction $\delta f^{(2)}$ is a few orders of magnitude smaller than the first-order correction $\delta f^{(1)}$. Transition arrays with two electrons in the 3d subshell, as well as those with a populated $4f$ subshell, exhibit the strongest corrections.

\begin{table}[!ht]
\begin{center}
\begin{tabular}{lrrrrrr}\hline\hline
Transition array & Nb. lines & $E_{CC'}$& $\delta E_{CC'}$ & $\sigma_{CC'}^2$ & $\delta f^{(1)}$ & $\delta f^{(2)}$\\\hline\hline
2s$^2$ 2p$^5$ - 2s$^2$ 2p$^4$ 3d$^1$ & 34 & 868.94729 & 15.4425 & 9.1149\,(6.1492) & 0.5331$\times 10^{-1}$ & 0.3621$\times 10^{-4}$\\
2s$^2$ 2p$^4$ - 2s$^2$ 2p$^3$ 3d$^1$ & 97 & 913.77475 & 17.0359 & 10.4790\,(6.3725) & 0.5593$\times 10^{-1}$ & 0.3765$\times 10^{-4}$\\
2s$^2$ 2p$^3$ - 2s$^2$ 2p$^2$ 3d$^1$ & 97 & 958.4424 & 18.6101 & 10.6906\,(6.6008) & 0.5825$\times 10^{-1}$ & 0.3491$\times 10^{-4}$\\
2s$^1$ 2p$^5$ - 2s$^1$ 2p$^4$ 3d$^1$ & 123 & 903.7576 & 16.99465 & 16.9305\,(6.3147) & 0.5641$\times 10^{-1}$ & 0.6219$\times 10^{-4}$\\
2s$^2$ 2p$^6$ - 2s$^2$ 2p$^5$ 3d$^1$ & 3 & 823.89045 & 13.8373 & 5.9308\,(5.9308) & 0.5039$\times 10^{-1}$ & 0.2621$\times 10^{-4}$\\
2s$^1$ 2p$^4$ - 2s$^1$ 2p$^3$ 3d$^1$ & 353 & 948.2068 & 18.5516 & 17.6586\,(6.5426) & 0.5869$\times 10^{-1}$ & 0.5892$\times 10^{-4}$\\
2s$^1$ 2p$^3$ - 2s$^1$ 2p$^2$ 3d$^1$ & 353 & 992.5311 & 20.0788 & 17.7337\,(6.7765) & 0.6069$\times 10^{-1}$ & 0.5400$\times 10^{-4}$\\
2s$^2$ 2p$^5$ 3d$^1$ - 2s$^2$ 2p$^4$ 3d$^2$ & 727 & 856.3464 & 14.5121 & 9.6348\,(6.1381) & 0.5084$\times 10^{-1}$ & 0.3942$\times 10^{-4}$\\
2s$^2$ 2p$^4$ 3d$^1$ - 2s$^2$ 2p$^3$ 3d$^2$ & 2190 & 900.8040 & 16.0956 & 11.0269\,(6.3588) & 0.5360$\times 10^{-1}$ & 0.4077$\times 10^{-4}$\\
2s$^1$ 2p$^6$ - 2s$^1$ 2p$^5$ 3d$^1$ & 11 & 859.1696 & 15.4164 & 15.4987\,(6.0924) & 0.5383$\times 10^{-1}$ & 0.6299$\times 10^{-4}$\\
2s$^2$ 2p$^5$ 3p$^1$ - 2s$^2$ 2p$^4$ 3p$^1$ 3d$^1$ & 860 & 857.8413 & 14.6418 & 9.8920\,(6.1185) & 0.5120$\times 10^{-1}$ & 0.4033$\times 10^{-4}$\\
2s$^2$ 2p$^4$ 3p$^1$ - 2s$^2$ 2p$^3$ 3p$^1$ 3d$^1$ & 2512 & 902.6094 & 16.2294 & 11.1892\,(6.3398) & 0.5394$\times 10^{-1}$ & 0.4120$\times 10^{-4}$\\
2s$^2$ 2p$^2$ - 2s$^2$ 2p$^1$ 3d$^1$ & 34 & 1002.9668 & 20.1574 & 9.7467\,(6.8340) & 0.6029$\times 10^{-1}$ & 0.2907$\times 10^{-4}$\\
2s$^2$ 2p$^3$ 3d$^1$ - 2s$^2$ 2p$^2$ 3d$^2$ & 2190 & 945.0847 & 17.6670 & 11.3168\,(6.5839) & 0.5608$\times 10^{-1}$ & 0.3801$\times 10^{-4}$\\
2s$^1$ 2p$^5$ 3d$^1$ - 2s$^1$ 2p$^4$ 3d$^2$ & 2722 & 890.6767 & 16.06315 & 17.3418\,(6.3004) & 0.5410$\times 10^{-1}$ & 0.6558$\times 10^{-4}$\\
2s$^2$ 2p$^6$ 3d$^1$ - 2s$^2$ 2p$^5$ 3d$^2$ & 60 & 811.7013 & 12.92455 & 6.5327\,(5.9221) & 0.4777$\times 10^{-1}$ & 0.2975$\times 10^{-4}$\\
2s$^1$ 2p$^4$ 3d$^1$ - 2s$^1$ 2p$^3$ 3d$^2$ & 8231 & 934.7501 & 17.6190 & 18.1093\,(6.5248) & 0.5655$\times 10^{-1}$ & 0.6218$\times 10^{-4}$\\
2s$^2$ 2p$^5$ 4f$^1$ - 2s$^2$ 2p$^4$ 4f$^1$ 3d$^1$ & 1707 & 866.9846 & 15.0867 & 9.5826\,(6.1536) & 0.5220$\times 10^{-1}$ & 0.3885$\times 10^{-4}$\\
2s$^2$ 2p$^3$ 3p$^1$ - 2s$^2$ 2p$^2$ 3p$^1$ 3d$^1$ & 2512 & 947.1816 & 17.8001 & 11.4084\,(6.5660) & 0.5638$\times 10^{-1}$ & 0.3815$\times 10^{-4}$\\
2s$^2$ 2p$^4$ 4f$^1$ - 2s$^2$ 2p$^3$ 4f$^1$ 3d$^1$ & 5332 & 911.7497 & 16.6596 & 10.9511\,(6.3772) & 0.5482$\times 10^{-1}$ & 0.3952$\times 10^{-4}$\\\hline\hline
\end{tabular}
\caption{Parameters of various 2p-3d transition arrays in an iron plasma at $\rho=$0.17 g/cm$^3$ and $T=$182 eV. The K shell is always full. The values inside parentheses in the fifth column are the spin-orbit contributions to the variance. $E_{CC'}$ and $\delta E_{CC'}$ are in eV, $\sigma_{CC'}^2$ is in eV$^2$.}\label{tab2}
\end{center}
\end{table}

\begin{figure}
\begin {center}
\includegraphics[scale=0.45]{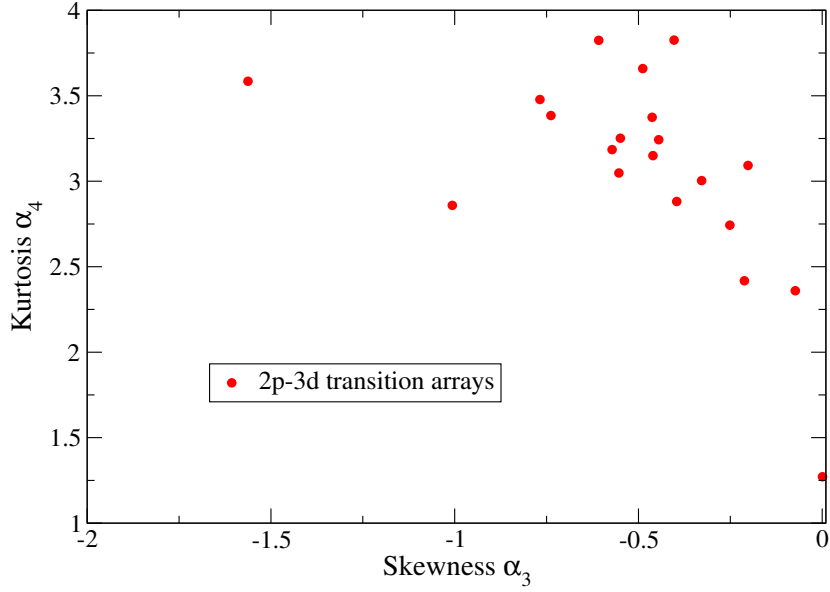}
\caption{Kurtosis ($\alpha_4$) versus skewness ($\alpha_3$) for the 2p-3d transition arrays displayed in table \ref{tab3}.}\label{fig1_a3a4}
\end{center}
\end{figure}

The reduced centered moment of order $n$ therefore reads:
\begin{equation*}
    \alpha_n=\frac{\mu_n}{\sigma^n}.
\end{equation*}
Figure \ref{fig1_a3a4} displays the kurtosis ($\alpha_4$) versus skewness ($\alpha_3$) for the 2p-3d transition arrays listed in table \ref{tab3}. The numerical values of the reduced centered moments up to $\alpha_6$ are provided in table \ref{tab3} of \ref{appA}. For the Gaussian assumption underlying the UTA formalism, one has $\alpha_3=\alpha_5=0$, $\alpha_4=3$ and $\alpha_6=15$. We observe that all the skewness values are negative, indicating that the transition arrays are asymmetric toward the lower energies (i.e., left-skewed). The average skewness is equal to $\bar{\alpha}_3=-0.50391$. The kurtosis values are all very close to 3, suggesting that the distributions are nearly Gaussian. Fourteen of the distributions are flatter than Gaussian (i.e., platykurtic), while six are sharper (i.e., mesokurtic). The average kurtosis value is $\bar{\alpha}_4$ = 3.08164. 

The values of $\delta f^{(3)}$, which require $\alpha_3$, are even smaller than the ones of $\delta f^{(2)}$. This is the reason why they are not indicated in table \ref{tab2}. They will also be discarded for the other examples of the present work.

\subsection{2p-4d transitions}\label{subsec33}

The parameters of various 2p-4d transition arrays in an iron plasma at $\rho=$0.17 g/cm$^3$ and $T=$182 eV are given in table \ref{tab4}. The reduced centered moments up to $\alpha_6$ of various 2p-3d transition arrays in an iron plasma at $\rho=$0.17 g/cm$^3$ and $T=$182 eV are given in table \ref{tab5} of \ref{appA}. Here as well, the transition arrays are selected based on the first ones encountered by the code, specifically the 2p-4d transitions from the initial configurations with the highest probability. The conclusions are similar to those drawn for the 2p-3d transitions, with the difference that in this case, the correction values are of the same order of magnitude for all the transitions considered, including those involving an electron in the 4f subshell. 

\begin{table}[!ht]
\begin{center}
\begin{tabular}{lrrrrrrr}\hline\hline
Transition array & Nb. lines & $E_{CC'}$ & $\delta E_{CC'}$ & $\sigma^2$ & $\delta f^{(1)}$ & $\delta f^{(2)}$\\\hline\hline
2s$^2$ 2p$^5$ - 2s$^2$ 2p$^4$ 4d$^1$ & 34 & 1084.4551 & 5.21226 & 10.1557\,(6.3149) & 0.1442$\times 10^{-1}$ & 0.2591$\times 10^{-4}$\\
2s$^2$ 2p$^4$ - 2s$^2$ 2p$^3$ 4d$^1$ & 97 & 1154.82365 & 5.6333 & 11.8698\,(6.5679) & 0.1463$\times 10^{-1}$ & 0.2670$\times 10^{-4}$\\
2s$^2$ 2p$^3$ - 2s$^2$ 2p$^2$ 4d$^1$ & 97 & 1226.4677 & 6.0430 & 12.1352\,(6.8281) & 0.1478$\times 10^{-1}$ & 0.2420$\times 10^{-4}$\\
2s$^1$ 2p$^5$ - 2s$^1$ 2p$^4$ 4d$^1$ & 123 & 1142.1725 & 5.6301 & 19.0389\,(6.5070) & 0.1479$\times 10^{-1}$ & 0.4378$\times 10^{-4}$\\
2s$^2$ 2p$^6$ - 2s$^2$ 2p$^5$ 4d$^1$ & 3 & 1015.3130 & 4.7796 & 6.0693\,(6.0693) & 0.1412$\times 10^{-1}$ & 0.1766$\times 10^{-4}$\\
2s$^1$ 2p$^4$ - 2s$^1$ 2p$^3$ 4d$^1$ & 353 & 1213.2595 & 6.0396 & 20.1190\,(6.7662) & 0.1493$\times 10^{-1}$ & 0.4100$\times 10^{-4}$\\
2s$^1$ 2p$^3$ - 2s$^1$ 2p$^2$ 4d$^1$ & 353 & 1285.6047 & 6.4386 & 20.3694\,(7.0334) & 0.1502$\times 10^{-1}$ & 0.3697$\times 10^{-4}$\\
2s$^2$ 2p$^5$ 3d$^1$ - 2s$^2$ 2p$^4$ 3d$^1$ 4d$^1$ & 1506 & 1057.55315 & 4.8874 & 10.7461\,(6.2933) & 0.1386$\times 10^{-1}$ & 0.2882$\times 10^{-4}$\\
2s$^2$ 2p$^4$ 3d$^1$ - 2$s^2$ 2p$^3$ 3d$^1$ 4d$^1$ & 4557 & 1126.8807 & 5.3176 & 12.4795\,(6.5423) & 0.1416$\times 10^{-1}$ & 0.2948$\times 10^{-4}$\\
2s$^1$ 2p$^6$ - 2s$^1$ 2p$^5$ 4d$^1$ & 11 & 1072.36585 & 5.2096 & 17.0557\,(6.2557) & 0.1457$\times 10^{-1}$ & 0.4449$\times 10^{-4}$\\
2s$^2$ 2p$^5$ 3p$^1$ - 2s$^2$ 2p$^4$ 3p$^1$ 4d$^1$ & 860 & 1060.6277 & 4.92485 & 10.9995\,(6.2748) & 0.1393$\times 10^{-1}$ & 0.2933$\times 10^{-4}$\\
2s$^2$ 2p$^4$ 3p$^1$ - 2s$^2$ 2p$^3$ 3p$^1$ 4d$^1$ & 2512 & 1130.42975 & 5.3626 & 12.6549\,(6.5245) & 0.1423$\times 10^{-1}$ & 0.2971$\times 10^{-4}$\\
2s$^2$ 2p$^2$ - 2s$^2$ 2p$^1$ 4d$^1$ & 34 & 1299.3645 & 6.4416 & 10.9255\,(7.0954) & 0.1487$\times 10^{-1}$ & 0.1941$\times 10^{-4}$\\
2s$^2$ 2p$^3$ 3d$^1$ - 2s$^2$ 2p$^2$ 3d$^1$ 4d$^1$ & 4557 & 1197.4715 & 5.7364 & 12.8339\,(6.7982) & 0.1437$\times 10^{-1}$ & 0.2685$\times 10^{-4}$\\
2s$^1$ 2p$^5$ 3d$^1$ - 2s$^1$ 2p$^4$ 3d$^1$ 4d$^1$ & 5655 & 1114.2810 & 5.31308 & 19.4178\,(6.4812) & 0.1430$\times 10^{-1}$ & 0.4692$\times 10^{-4}$\\
2s$^2$ 2p$^6$ 3d$^1$ - 2s$^2$ 2p$^5$ 3d$^1$ 4d$^1$ & 123 & 989.5061 & 4.44602 & 6.8333\,(6.0512) & 0.1348$\times 10^{-1}$ & 0.2094$\times 10^{-4}$\\
2s$^1$ 2p$^4$ 3d$^1$ - 2s$^1$ 2p$^3$ 3d$^1$ 4d$^1$ & 17171 & 1184.3303 & 5.7310 & 20.5389\,(6.7358) & 0.1452$\times 10^{-1}$ & 0.4393$\times 10^{-4}$\\
2s$^2$ 2p$^5$ 4f$^1$ - 2s$^2$ 2p$^4$ 4f$^1$ 4d$^1$ & 1707 & 1076.3794 & 5.0885 & 10.4792\,(6.3160) & 0.1418$\times 10^{-1}$ & 0.2713$\times 10^{-4}$\\
2s$^2$ 2p$^3$ 3p$^1$ - 2s$^2$ 2p$^2$ 3p$^1$ 4d$^1$ & 2512 & 1201.4654 & 5.7887 & 12.9546\,(6.7814) & 0.1445$\times 10^{-1}$ & 0.2692$\times 10^{-4}$\\
2s$^2$ 2p$^4$ 4f$^1$ - 2s$^2$ 2p$^3$ 4f$^1$ 4d$^1$ & 5332 & 1146.2591 & 5.51695 & 12.1735\,(6.5686) & 0.1444$\times 10^{-1}$ & 0.2780$\times 10^{-4}$\\\hline\hline
\end{tabular}
\caption{Parameters of various 2p-4d transition arrays in an iron plasma at $\rho=$0.17 g/cm$^3$ and $T=$182 eV. The K shell is always full. The values inside parentheses in the fifth column are the spin-orbit contributions to the variance. $E_{CC'}$ and $\delta E_{CC'}$ are in eV, $\sigma^2$ is in eV$^2$.}\label{tab4}
\end{center}
\end{table}

\begin{figure}
\begin {center}
\includegraphics[scale=0.45]{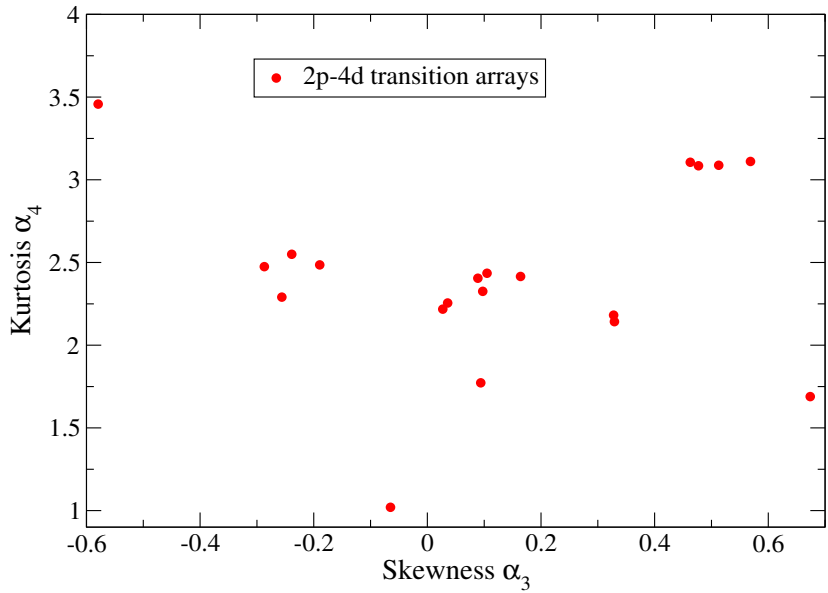}
\caption{Kurtosis ($\alpha_4$) versus skewness ($\alpha_3$) for the 2p-4d transition arrays reported in table \ref{tab5}.}\label{fig2_a3a4}
\end{center}
\end{figure}

Figure \ref{fig2_a3a4} represents the kurtosis versus skewness for the 2p-4d transition arrays of table \ref{tab5}. The numerical values of the reduced centered moments up to $\alpha_6$ of various 2p-4d transition arrays in an iron plasma at $\rho=$0.17 g/cm$^3$ and $T=$182 eV are given in table \ref{tab5} of \ref{appA}. Six arrays are left-skewed, and 14 are right-skewed. The kurtosis values are all close to 3, indicating that the distributions are nearly Gaussian. Five of them exhibit sharper peaks than the Gaussian (i.e., mesokurtic), while 15 are flatter (i.e., platykurtic). The average kurtosis value is 2.4253.

\section{Iron plasma at \texorpdfstring{$\rho=$0.01 g/cm$^3$ and $T=$22 eV}{rho=0.01 g/cm3 and T=22 eV}}\label{sec4}

\subsection{Average atomic structure}\label{subsec41}

Let us now investigate an iron plasma at $\rho=0.01$ g/cm$^3$ and $T=$22 eV. The average energies and electron populations of the subshells are provided in table \ref{tab6}. By ``average'', we mean that they are averaged over all the configurations generated by the code (the subshell energies vary from a configuration to another, and the populations are of course natural numbers). Such conditions refer to laser opacity measurements carried out over the past few decades and are of interest for envelopes of variable stars such as $\beta$-Cephei-type stars. The latter are pulsating stars of masses between $\approx 8$ and 25 solar masses. Their pulsations are low-order pressure (p) and gravity (g) modes with periods typically of $\approx 0.5$ to $8$ hours. Since part of these $\beta$-Cephei modes present a mixed p- and g- character, it turns out that they are privileged targets to test physical processes at the boundary of the convective core and radiative envelope with asteroseismology \cite{Salmon2022}. Finding an instability mechanism to explain the pulsations in $\beta$-Cephei stars and other B-type variables has long been a significant challenge for the theory of stellar structure and evolution \cite{Moskalik1992}. In 1978, Stellingwerf observed that a bump in opacity near $T = 150,000$ K, caused by the He II ionization edge, had a destabilizing effect on the fundamental radial mode \cite{Stellingwerf1978}. However, this feature was not large enough to induce instability in any radial or non-radial mode. A significant enhancement of the bump was required to counteract the damping effects occurring elsewhere in the star. Subsequently, Simon pointed out that increasing the opacity of heavy elements, which would address the long-standing discrepancy between predicted and observed period ratios in Cepheids, would simultaneously lead to the desired enhancement of the bump \cite{Simon1982}.

\begin{table}[!ht]
\begin{subtable}[c]{0.5\textwidth}
\begin{tabular}{rrrr}\hline\hline
$n$ & $\ell$ & Energy & Population\\\hline\hline
1 & 0 & -0.7124$\times 10^{4}$ & 0.2000$\times 10^1$\\
2 & 0 & -0.9561$\times 10^{3}$ & 0.2000$\times 10^1$\\
2 & 1 & -0.8365$\times 10^{3}$ & 0.6000$\times 10^1$\\
3 & 0 & -0.2136$\times 10^{3}$ & 0.1892$\times 10^1$\\
3 & 1 & -0.17855$\times 10^{3}$ & 0.4682$\times 10^1$\\
3 & 2 & -0.1231$\times 10^{3}$ & 0.2225$\times 10^1$\\
4 & 0 & -0.7270$\times 10^{2}$ & 0.5621$\times 10^{-1}$\\
4 & 1 & -0.6178$\times 10^{2}$ & 0.1038\\
4 & 2 & -0.4542$\times 10^{2}$ & 0.8302$\times 10^{-1}$\\
4 & 3 & -0.3254$\times 10^{2}$ & 0.6495$\times 10^{-1}$\\
5 & 0 & -0.3243$\times 10^{2}$ & 0.9234$\times 10^{-2}$\\
5 & 1 & -0.2802$\times 10^{2}$ & 0.2269$\times 10^{-1}$\\
5 & 2 & -0.2128$\times 10^{2}$ & 0.2786$\times 10^{-1}$\\
5 & 3 & -0.1583$\times 10^{2}$ & 0.3046$\times 10^{-1}$\\
5 & 4 & -0.1384$\times 10^{2}$ & 0.3579$\times 10^{-1}$\\
6 & 0 & -0.1567$\times 10^{2}$ & 0.4322$\times 10^{-1}$\\\hline\hline
\end{tabular}
\end{subtable}
\begin{subtable}[c]{0.5\textwidth}
\begin{tabular}{rrrr}\hline\hline
$n$ & $\ell$ & Energy & Population\\\hline\hline
6 & 1 & -0.1349$\times 10^{2}$ & 0.11745$\times 10^{-1}$\\
6 & 2 & -0.10105$\times 10^{2}$ & 0.16785$\times 10^{-1}$\\
6 & 3 & -0.7301$\times 10^{1}$ & 0.2069$\times 10^{-1}$\\
6 & 4 & -0.6184$\times 10^{1}$ & 0.2529$\times 10^{-1}$\\
6 & 5 & -0.5767$\times 10^{1}$ & 0.3033$\times 10^{-1}$\\
7 & 0 & -0.7261$\times 10^{1}$ & 0.29505$\times 10^{-2}$\\
7 & 1 & -0.6066$\times 10^{1}$ & 0.8384$\times 10^{-2}$\\
7 & 2 & -0.4191$\times 10^{1}$ & 0.1283$\times 10^{-1}$\\
7 & 3 & -0.2622$\times 10^{1}$ & 0.1673$\times 10^{-1}$\\
7 & 4 & -0.1944$\times 10^{1}$ & 0.2086$\times 10^{-1}$\\
7 & 5 & -0.16255$\times 10^{1}$ & 0.2513$\times 10^{-1}$\\
7 & 6 & -0.1364$\times 10^{1}$ & 0.2935$\times 10^{-1}$\\
8 & 0 & -0.2679$\times 10^{1}$ & 0.23965$\times 10^{-2}$\\
8 & 1 & -0.2004$\times 10^{1}$ & 0.6972$\times 10^{-2}$\\
8 & 2 & -0.9654$\times 10^{0}$ & 0.11085$\times 10^{-1}$\\
8 & 3 & -0.13225$\times 10^{0}$ & 0.1494$\times 10^{-1}$\\\hline\hline
\end{tabular}
\end{subtable}
\caption{Energies (in eV) and populations of the different subshells, in an iron plasma at $\rho=$0.01 g/cm$^3$ and $T=$22 eV.}\label{tab6}
\end{table}

\subsection{3p-3d transitions}\label{subsec42}

The parameters of various 3p-3d transition arrays in an iron plasma at $\rho=$0.01 g/cm$^3$ and $T=$22 eV are given in table \ref{tab7}. The K shell is always full. The spin-orbit contribution to the variance is much smaller than for the 2p-3d and 2p-4d transitions. The relative contribution of the UTA shift $\delta E_{CC'}$ to the energy $E_{CC'}$ is also more pronounced. There are only two orders of magnitude between the first- and second-order corrections. Both are much larger than for the T=182 eV, $\rho$=0.17 g/cm$^3$ case. The corrections are almost constant from a transition array to another. 

\begin{table}[!ht]
\begin{center}
\begin{tabular}{lrrrrrrr}\hline\hline
Transition array & Nb. lines & $E_{CC'}$ & $\delta E_{CC'}$ & $\sigma^2$ & $\delta f^{(1)}$ & $\delta f^{(2)}$\\\hline\hline
3s$^2$ 3p$^5$ 3d$^2$ - 3s$^2$ 3p$^4$ 3d$^3$ & 22481 & 69.7124 & 19.8400 & 6.5361\,(0.7890) & 0.8943 & 0.4959$\times 10^{-2}$\\
3s$^2$ 3p$^5$ 3d$^1$ - 3s$^2$ 3p$^4$ 3d$^2$ & 727 & 71.0636 & 20.2510 & 5.5443\,(0.8258) & 0.8549 & 0.3294$\times 10^{-2}$\\
3s$^2$ 3p$^5$ 3d$^3$ - 3s$^2$ 3p$^4$ 3d$^4$ & 22329 & 68.0960 & 19.3592 & 7.0085\,(0.7556) & 0.8529 & 0.4534$\times 10^{-2}$\\
3s$^2$ 3p$^4$ 3d$^2$ - 3s$^2$ 3p$^3$ 3d$^3$ & 18237 & 66.5103 & 20.2495 & 7.0965\,(0.8241)& 0.9134 & 0.4813$\times 10^{-2}$\\
3s$^2$ 3p$^6$ 3d$^2$ - 3s$^2$ 3p$^5$ 3d$^3$ & 466 & 72.49535 & 19.3510 & 5.4932\,(0.7556) & 0.8008 & 0.3136$\times 10^{-2}$\\
3s$^2$ 3p$^4$ 3d$^1$ - 3s$^2$ 3p$^3$ 3d$^2$ & 2190 & 67.5380 & 20.6003 & 6.1618\,(0.8629) & 0.9151 & 0.4053$\times 10^{-2}$\\
3s$^2$ 3p$^4$ 3d$^3$ - 3s$^2$ 3p$^3$ 3d$^4$ & 69501 & 65.2160 & 19.84165 & 7.5583\,(0.7882) & 0.9127 & 0.5331$\times 10^{-2}$\\
3s$^2$ 3p$^6$ 3d$^1$ - 3s$^2$ 3p$^5$ 3d$^2$ & 60 & 74.20605 & 19.8394 & 4.2840\,(0.7900) & 0.8021 & 0.2334$\times 10^{-2}$\\
3s$^2$ 3p$^6$ 3d$^3$ - 3s$^2$ 3p$^5$ 3d$^4$ & 1718 & 70.5063 & 18.7726 & 6.0398\,(0.7251) & 0.7988 & 0.3645$\times 10^{-2}$\\
3s$^2$ 3p$^5$ 3d$^4$ - 3s$^2$ 3p$^4$ 3d$^5$ & 42579 & 66.5103 & 18.7956 & 7.0952\,(0.7261) & 0.8478 & 0.4812$\times 10^{-2}$\\
3s$^2$ 3p$^5$ 3d$^0$ - 3s$^2$ 3p$^4$ 3d$^1$ & 34 & 72.1263 & 20.6020 & 3.5894\,(0.8654) & 0.8569 & 0.2070$\times 10^{-2}$\\
3s$^2$ 3p$^4$ 3d$^4$ - 3s$^2$ 3p$^3$ 3d$^5$ & 133102 & 63.6929 & 19.3684 & 7.6572\,(0.7557) & 0.9123 & 0.5663$\times 10^{-2}$\\
3s$^2$ 3p$^3$ 3d$^2$ - 3s$^2$ 3p$^2$ 3d$^3$ & 18237 & 62.9411 & 20.5990 & 7.2271\,(0.8607) & 0.9818 & 0.5473$\times 10^{-2}$\\
3s$^2$ 3p$^6$ 3d$^4$ - 3s$^2$ 3p$^5$ 3d$^5$ & 3245 & 68.2741 & 18.0823 & 6.1421\,(0.6993) & 0.7945 & 0.3953$\times 10^{-2}$\\
3s$^2$ 3p$^3$ 3d$^1$ - 3s$^2$ 3p$^2$ 3d$^2$ & 2190 & 63.67995 & 20.9009 & 6.2644\,(0.9013) & 0.9847 & 0.4634$\times 10^{-2}$\\
3s$^2$ 3p$^3$ 3d$^3$ - 3s$^2$ 3p$^2$ 3d$^4$ & 69501 & 61.9450 & 20.2487 & 7.7180\,(0.8226) & 0.9806 & 0.6034$\times 10^{-2}$\\
3s$^2$ 3p$^4$ 3d$^0$ - 3s$^2$ 3p$^3$ 3d$^1$ & 97 & 68.2883 & 20.9009 & 4.4376\,(0.9042) & 0.9182 & 0.2855$\times 10^{-2}$\\
3s$^2$ 3p$^6$ 3d$^0$ - 3s$^2$ 3p$^5$ 3d$^1$ & 3 & 75.6151 & 20.2533 & 0.8276\,(0.8276) & 0.8035 & 0.4342$\times 10^{-3}$\\
3s$^2$ 3p$^5$ 3d$^5$ - 3s$^2$ 3p$^4$ 3d$^6$ & 42579 & 64.1982 & 18.13166 & 6.8516\,(0.7014) & 0.8473 & 0.4987$\times 10^{-2}$\\
3s$^1$ 3p$^5$ 3d$^2$ - 3s$^1$ 3p$^4$ 3d$^3$ & 22481 & 67.9249 & 20.2491 & 7.6264\,(0.8244) & 0.8943 & 0.4959$\times 10^{-2}$\\\hline\hline
\end{tabular}
\caption{Parameters of various 3p-3d transition arrays in an iron plasma at $\rho=$0.01 g/cm$^3$ and $T=$22 eV. The K shell is always full. The values inside parentheses in the fifth column are the spin-orbit contributions to the variance. $E_{CC'}$ and $\delta E_{CC'}$ are in eV, $\sigma^2$ is in eV$^2$}\label{tab7}
\end{center}
\end{table}

\begin{figure}
\begin {center}
\includegraphics[scale=0.45]{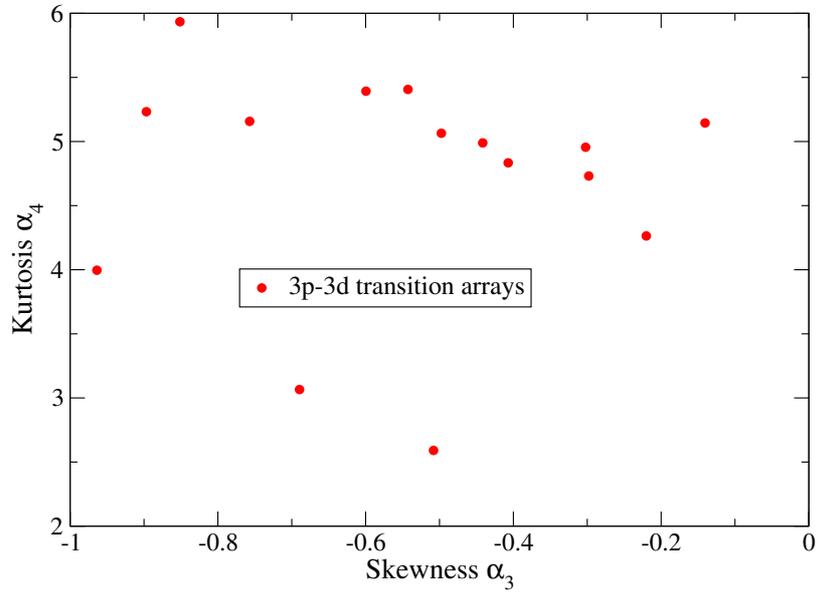}
\caption{Kurtosis ($\alpha_4$) versus skewness ($\alpha_3$) for the $3p-3d$ transition arrays displayed in table \ref{tab8}.}\label{fig3_a3a4}
\end{center}
\end{figure}

In the case of the 3p-3d transition arrays, the kurtosis is plotted versus skewness in figure \ref{fig3_a3a4}. The reduced centered moments of various 3p-3d transition arrays are given in table \ref{tab8} of \ref{appA}. We can see that all the skewness $\alpha_3$ values are negative, which means that the transition arrays are all asymmetric to the lower energies (i.e., left-skewed). The average skewness value is equal to $\alpha_3$ = -1.95355. Excluding the pathological very sharp array 3s$^2$ 3p$^6$ 3d$^0$ - 3s$^2$ 3p$^5$ 3d$^1$ (having only 3 lines and thus not suitable for a statistical modeling), the kurtosis are all very close to 3, which means that the distributions are close to Gaussian ones. Nineteen of them are sharper (i.e. mesokurtic) and only one (3s$^2$ 3p$^6$ 3d$^4$ - 3s$^2$ 3p$^5$ 3d$^5$) is flatter (platykurtic). We find the average kurtosis value $\bar{\alpha_4}=33.0148$.

\subsection{3s-3p transitions}\label{subsec43}

The parameters for various 3s-3p transition arrays in an iron plasma at $\rho = 0.01 , \text{g/cm}^3$ and $T = 22 \text{eV}$ are provided in table \ref{tab9}. Regarding the corrections, the results closely resemble those from the previous 3p-3d transitions, and the same conclusions apply.

\begin{table}[!ht]
\begin{center}
\begin{tabular}{lrrrrrrr}\hline\hline
Transition array & Nb. lines & $E_{CC'}$ & $\delta E_{CC'}$ & $\sigma^2$ & $\delta f^{(1)}$ & $\delta f^{(2)}$\\\hline\hline
3s$^2$ 3p$^5$ 3d$^2$ - 3s$^1$ 3p$^6$ 3d$^2$ & 401 & 42.8720 & 10.5856 & 5.6916\,(0.9024) & 0.7407 & 0.9290$\times 10^{-2}$\\
3s$^2$ 3p$^5$ 3d$^1$ - 3s$^1$ 3p$^6$ 3d$^1$ & 36 & 43.5978 & 10.7833 & 4.4138\,(0.9500) & 0.7420 & 0.6966$\times 10^{-2}$\\
3s$^2$ 3p$^5$ 3d$^3$ - 3s$^1$ 3p$^6$ 3d$^3$ & 2082 & 42.1288 & 10.3918 & 6.3302\,(0.8588) & 0.7400 & 0.1070$\times 10^{-1}$\\
3s$^2$ 3p$^4$ 3d$^2$ - 3s$^1$ 3p$^5$ 3d$^2$ & 5015 & 43.1140 & 10.7764 & 7.5716\,(0.9475) & 0.7499 & 0.1222$\times 10^{-1}$\\
3s$^2$ 3p$^4$ 3d$^1$ - 3s$^1$ 3p$^5$ 3d$^1$ & 413 & 43.7942 & 10.9721 & 6.6700\,(0.9975) & 0.7516 & 0.1043$\times 10^{-1}$\\
3s$^2$ 3p$^4$ 3d$^3$ - 3s$^1$ 3p$^5$ 3d$^3$ & 26903 & 42.4040 & 10.5829 & 8.0576\,(0.9010) & 0.7487 & 0.1344$\times 10^{-1}$\\
3s$^2$ 3p$^5$ 3d$^4$ - 3s$^1$ 3p$^6$ 3d$^4$ & 5424 & 41.3890 & 10.2055 & 6.5526\,(0.8198) & 0.7397 & 0.1148$\times 10^{-1}$\\
3s$^2$ 3p$^5$ 3d$^0$ - 3s$^1$ 3p$^6$ 3d$^0$ & 2 & 44.2898 & 10.9824 & 1.0009\,(1.0009) & 0.7439 & 0.1531$\times 10^{-2}$\\
3s$^2$ 3p$^4$ 3d$^4$ - 3s$^1$ 3p$^5$ 3d$^4$ & 71106 & 41.6809 & 10.3940 & 8.2267\,(0.8587) & 0.7481 & 0.1421$\times 10^{-1}$\\
3s$^2$ 3p$^3$ 3d$^2$ - 3s$^1$ 3p$^4$ 3d$^2$ & 15345 & 43.3001 & 10.9625 & 8.4562\,(0.9942) & 0.7595 & 0.1353$\times 10^{-1}$\\
3s$^2$ 3p$^3$ 3d$^1$ - 3s$^1$ 3p$^4$ 3d$^1$ & 1221 & 43.9285 & 11.1558 & 7.6444\,(1.0464) & 0.7619 & 0.1188$\times 10^{-1}$\\
3s$^2$ 3p$^3$ 3d$^3$ - 3s$^1$ 3p$^4$ 3d$^3$ & 85323 & 42.63175 & 10.7702 & 8.9045\,(0.9451) & 0.7579 & 0.1470$\times 10^{-1}$\\
3s$^2$ 3p$^4$ 3d$^0$ - 3s$^1$ 3p$^5$ 3d$^0$ & 14 & 44.4308 & 11.1682 & 5.1073\,(1.0506) & 0.7541 & 0.7761$\times 10^{-2}$\\
3s$^2$ 3p$^5$ 3d$^5$ - 2s$^2$ 2p$^6$ 3s$^1$ 3p$^6$ 3d$^5$ & 7426 & 40.6791 & 10.0320 & 6.4434\,(0.7862) & 0.7398 & 0.1168$\times 10^{-1}$\\
3s$^1$ 3p$^5$ 3d$^2$ - 2s$^2$ 2p$^6$ 3s$^0$ 3p$^6$ 3d$^2$ & 401 & 36.09945 & 10.7771 & 5.8233\,(0.9477) &	0.8956 & 0.1341$\times 10^{-1}$\\\hline\hline
\end{tabular}
\caption{Parameters of various 3s-3p transition arrays in an iron plasma at $\rho=$0.01 g/cm$^3$ and $T=$22 eV. The K shell is always full. The values inside parentheses in the fifth column are the spin-orbit contributions to the variance. $E_{CC'}$ and $\delta E_{CC'}$ are in eV, $\sigma^2$ is in eV$^2$}\label{tab9}
\end{center}
\end{table}

\begin{figure}
\begin {center}
\includegraphics[scale=0.45]{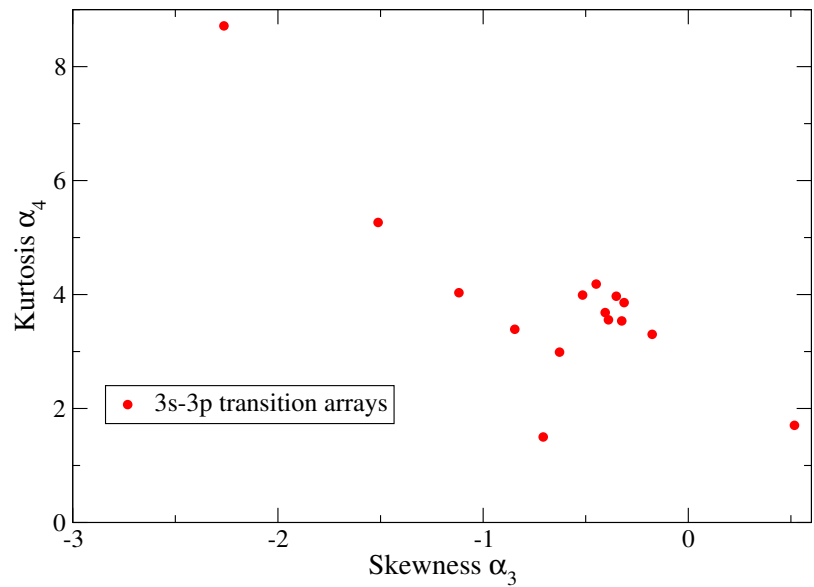}
\caption{Kurtosis versus skewness for the 3s-3p transition arrays displayed in table \ref{tab10}.}\label{fig4_a3a4}
\end{center}
\end{figure}

In figure \ref{fig4_a3a4} we represent the kurtosis versus the skewness for 3s-3p transition arrays. The numerical values of the reduced centered moments $\alpha_3$, $\alpha_4$, $\alpha_5$ and $\alpha_6$ of various 3s-3p are given in table \ref{tab10} of \ref{appA}. We can see that almost all the $\alpha_3$ values are negative (except for: 3s$^2$ 3p$^4$ 3d$^0$ - 3s$^1$ 3p$^5$ 3d$^0$), which means that the transition arrays are all asymmetric to the lower energies (i.e., left-skewed). The average skewness $\bar{\alpha}_3$ = -0.631947. The kurtosis are all rather close to 3, which means that the distributions are close to Gaussian ones. Twelve of them are sharper (i.e., mesokurtic) and 3 of them are flatter (platykurtic). One finds $\bar{\alpha}_4=3.84537$.

\subsection{Effect on radiative opacity}\label{subsec44}

\begin{figure}
\begin {center}
\includegraphics[scale=0.45]{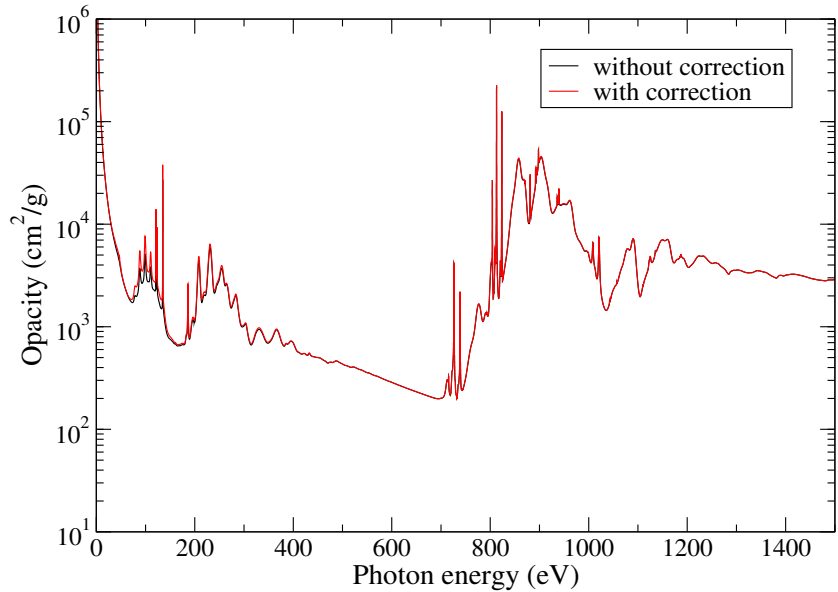}
\caption{Opacity of an iron plasma at $\rho=$0.17 g/cm$^3$ and $T=$182 eV with and without the corrections.}\label{fig1}
\end{center}
\end{figure}

\begin{figure}
\begin {center}
\includegraphics[scale=0.45]{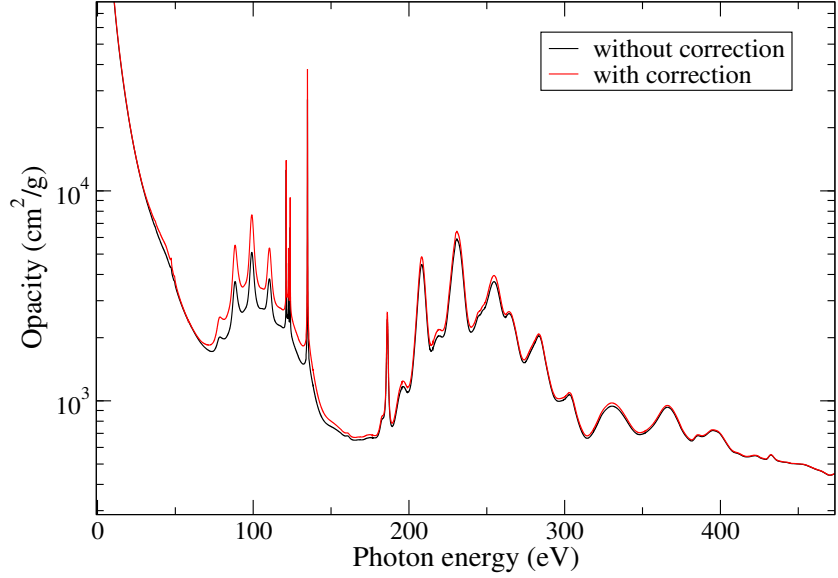}
\caption{Opacity of an iron plasma at $\rho=$0.17 g/cm$^3$ and $T=$182 eV with and without the corrections. Zoom on the preceding figure over the XUV range.}\label{fig2}
\end{center}
\end{figure}

\begin{figure}
\begin {center}
\includegraphics[scale=0.45]{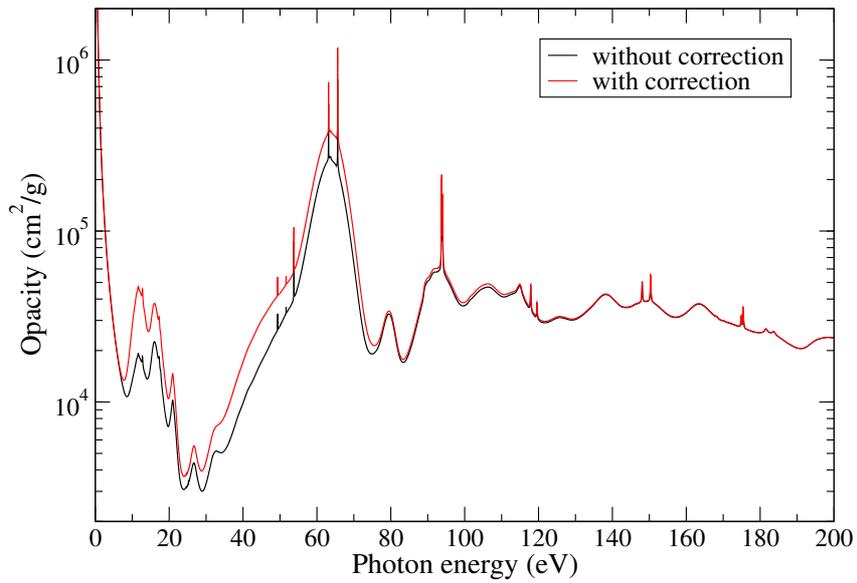}
\caption{Opacity of an iron plasma at $\rho=$0.01 g/cm$^3$ and $T=$22 eV with and without the corrections.}\label{fig3}
\end{center}
\end{figure}

Figures \ref{fig1} and \ref{fig2} represent the opacity of an iron plasma at $\rho=$0.17 g/cm$^3$ and $T=$182 eV. We can see that the effect of the corrections on spectral radiative opacity is significant. In the lower density plasma, these corrections play a noticeable role (see figure \ref{fig3}). As expected, the effect is more pronounced in this plasma (i.e., for $\Delta n$=0 transitions such as 3s-3p and 3p-3d) than in the higher density plasma, where the opacity is dominated by L-shell 2p-$n$d (mostly with $n$=3 and 4) transitions. The most noticeable differences in Fig. \ref{fig1} occur at $h\nu\approx 100$ eV, which corresponds to the 3-3 transitions as well. The Rosseland mean opacity, defined as
\begin{equation*}
    \kappa_R(u)=\left(\int_0^{\infty}\frac{W_R(u)}{\kappa(u)}\,\mathrm{d}u\right)^{-1},
\end{equation*}
where $\kappa(u)$ is the spectral opacity and
\begin{equation*}
    W_R(u)=\frac{15}{4\pi^4}\frac{u^4\,e^{-u}}{(1-e^{-u})^2}
\end{equation*}
the Rosseland weighting function, which is proportional to the derivative of the Planck distribution with respect to the temperature. The Rosseland mean is thus a weighted harmonic mean of the spectral (monochromatic) opacity and thus very sensitive to gaps in the spectrum. The Rosseland mean opacity is increased from 730.986 to 733.582 cm$^2$/g in Figs. \ref{fig1} and \ref{fig2} when the corrections are included. For Fig. \ref{fig3}, the Rosseland mean is increased from 7909.54 to 7913.91 cm$^2$/g when the corrections are included.

\section{Configuration-averaged radiative rates and detailed balance}\label{sec5}

\subsection{Detailed balance and recovering the LTE limit}\label{subsec53}

In the following, $N_C$ and $N_{C'}$ are the populations of configurations $C$ and $C'$, respectively. The emission and absorption coefficients read
\begin{equation*}
    j(\nu)=\frac{E_{CC'}}{4\pi}N_{C'}A_{C'C}\Phi_{C'C}^{\mathrm{sp}}(\nu)
\end{equation*}
and
\begin{equation*}
    \kappa(\nu)=\frac{E_{CC'}}{4\pi}\left(N_CB_{CC'}\Phi_{CC'}^{\mathrm{abs}}(\nu)-N_{C'}B_{C'C}\Phi_{C'C}^{\mathrm{st}}(\nu)\right), 
\end{equation*}
respectively, where \sout{$\bar{\nu}$ ....} $\Phi^{\mathrm{sp}}$, $\Phi^{\mathrm{abs}}$ and $\Phi^{\mathrm{st}}$ are the spontaneous-emission, absorption and stimulated-emission configuration ``average'' line profiles. The source function

\begin{figure}[!ht]
\begin {center}
\includegraphics[scale=0.4]{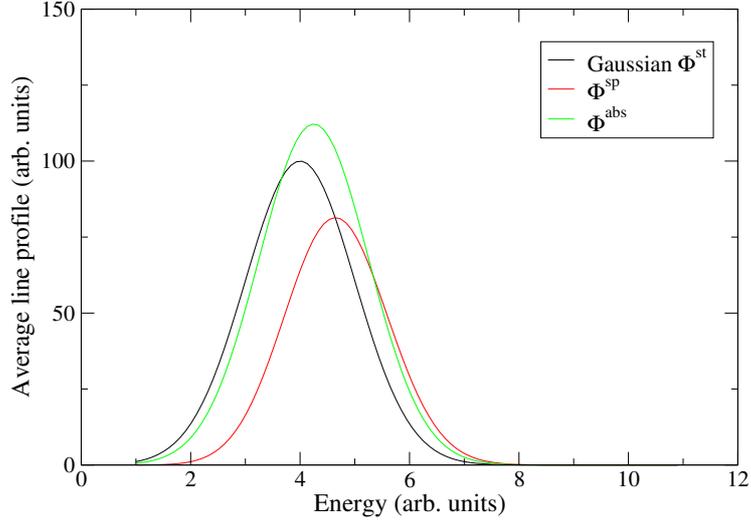}
\caption{Scaled configuration-averaged line profiles in the case of a Gaussian stimulated-emission profile. The represented functions are $100\,e^{-(x-4)^2/2}$, $x^3\,e^{-(x-4)^2/2}$ and $40\, e^{x/4}\,e^{-(x-4)^2/2}$.}\label{figna}
\end{center}
\end{figure}

\begin{figure}[!ht]
\begin {center}
\includegraphics[scale=0.4]{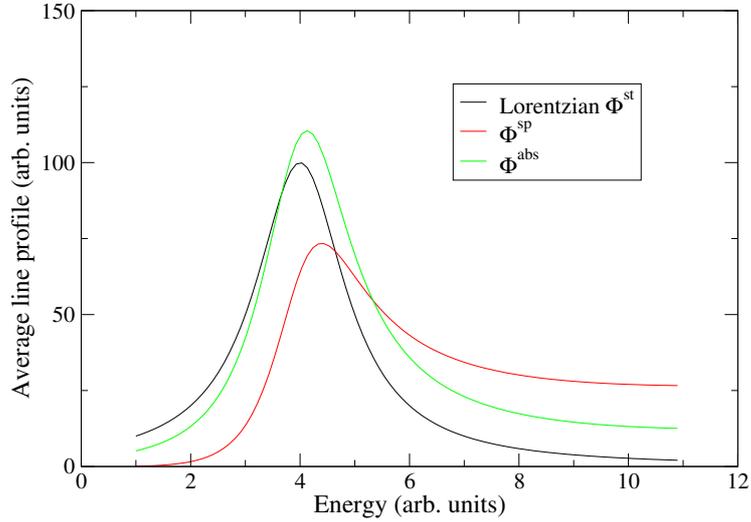}
\caption{Scaled configuration-averaged line profiles in the case of a Lorentzian stimulated-emission profile. The represented functions are $100/\left[1+(x-4)^2\right]$, $x^3/\left[1+(x-4)^2\right]$ and $40\, e^{x/4}/\left[1+(x-4)^2\right]$.}\label{fignb}
\end{center}
\end{figure}

\begin{equation}\label{source-function}
    S_{CC'}(\nu)=\frac{j_{CC'}(\nu)}{\kappa_{C'C}(\nu)}
\end{equation}
is therefore (using Eq. (\ref{avecmu3})):
\begin{equation*}
    S_{CC'}(\nu)\approx\frac{2}{h^2c^2}\frac{\mu_3\,e^{-\beta\delta E_{CC'}}}{\displaystyle\frac{N_C}{N_{C'}}\frac{g_{C'}}{g_C}\frac{\Phi_{C'C}^{\mathrm{abs}}(\nu)}{\Phi_{C'C}^{\mathrm{st}}(\nu)}-1}\frac{\Phi_{C'C}^{\mathrm{sp}}(\nu)}{\Phi_{C'C}^{\mathrm{st}}(\nu)}.
\end{equation*}
Thus, Kirhhoff's law is violated under LTE conditions. To address this issue, some authors have proposed introducing a heuristic correction (dependent on temperature and frequency) to the population of the upper states of a transition \cite{Makhrov1990}, modifying the emission and absorption profiles \cite{Busquet2003}, or even allowing the Einstein coefficients to vary with density and temperature \cite{Zemtsov1993}. Following Ref. \cite{Busquet2003}, we adopt an {\it ansatz} for the average line profiles, that preserves the Einstein relations. Consequently, to recover the Planckian distribution,
\begin{equation}\label{pla}
    S_{CC'}(\nu)\approx\frac{2h\nu^3}{c^2}\frac{1}{e^{\beta h\nu}-1},
\end{equation}
one can set
\begin{equation*}
    \frac{2}{h^2c^2}\,\mu_3\,e^{-\beta\delta E_{CC'}}\frac{\Phi_{C'C}^{\mathrm{sp}}(\nu)}{\Phi_{C'C}^{\mathrm{st}}(\nu)}=\frac{2h\nu^3}{c^2}
\end{equation*}
and
\begin{equation*}
    \frac{N_C}{N_{C'}}\frac{g_{C'}}{g_C}\frac{\Phi_{C'C}^{\mathrm{abs}}(\nu)}{\Phi_{C'C}^{\mathrm{st}}(\nu)}=e^{\beta h\nu}.
\end{equation*}

\subsection{Detailed balance with the corrections}\label{subsec54}

\begin{figure}[!ht]
\begin {center}
\includegraphics[scale=0.4]{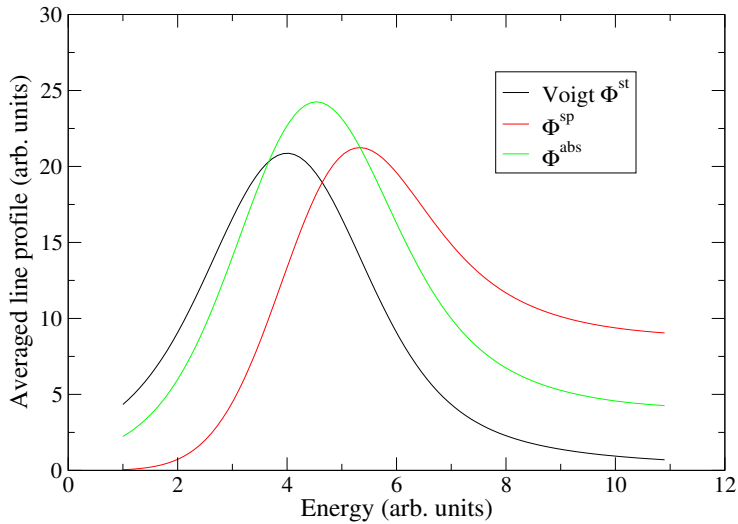}
\caption{Scaled configuration-averaged line profiles in the case of a Voigt stimulated-emission profile. The represented functions are $100\,\mathscr{V}(1,1,x-4)$, $x^3\,\mathscr{V}(1,1,x-4)$ and $40\,e^{x/4}\,\mathscr{V}(1,1,x-4)$, where the Voigt probability density function is given in Eq. (\ref{voi}).}\label{fignc}
\end{center}
\end{figure}

Accounting for the corrections requires to replace $A_{CC'}$ by
\begin{equation*}
    A_{CC'}^{\mathrm{corr}}=A_{CC'}\left(1+3\frac{\delta_{CC'}}{E_{CC'}}+3\frac{\sigma_{CC'}^2}{E_{CC'}^2}\right).
\end{equation*}
The source function (Eq. (\ref{source-function})) is therefore
\begin{equation*}
    S_{CC'}(\nu)\approx\frac{2}{h^2c^2}\frac{\mu_3\,e^{-\beta\delta_{CC'}}\displaystyle\left(1+3\frac{\delta E_{CC'}}{E_{CC'}}+3\frac{\sigma_{CC'}^2}{E_{CC'}^2}\right)}{\displaystyle\frac{N_C}{N_{C'}}\frac{g_{C'}}{g_C}\frac{\Phi_{C'C}^{\mathrm{abs}}(\nu)}{\Phi_{C'C}^{\mathrm{st}}(\nu)}-1}\frac{\Phi_{C'C}^{\mathrm{sp}}(\nu)}{\Phi_{C'C}^{\mathrm{st}}(\nu)}.
\end{equation*}
Thus, in order to recover the Planckian distribution (Eq. (\ref{pla})), we suggest setting:
\begin{equation*}
    \frac{2}{h^2c^2}\mu_3e^{-\beta\delta E_{CC'}}\left(1+3\frac{\delta E_{CC'}}{E_{CC'}}+3\frac{\sigma_{CC'}^2}{E_{CC'}^2}\right)\frac{\Phi_{C'C}^{\mathrm{sp}}(\nu)}{\Phi_{C'C}^{\mathrm{st}}(\nu)}=\frac{2h\nu^3}{c^2}
\end{equation*}
and
\begin{equation*}
    \frac{N_C}{N_{C'}}\frac{g_{C'}}{g_C}\frac{\Phi_{C'C}^{\mathrm{abs}}(\nu)}{\Phi_{C'C}^{\mathrm{st}}(\nu)}=e^{\beta h\nu}.
\end{equation*}
Given one profile, the two others can be obtained from the two relations above. For example, if $\Phi_{C'C}^{\mathrm{st}}$ is represented by a Gaussian shape, we easily obtain $\Phi_{C'C}^{\mathrm{sp}}$ and $\Phi_{C'C}^{\mathrm{abs}}$ (see Fig. \ref{figna}). Artificial scaling factors have been used to mimic the different configuration-dependent coefficients entering the formula. Figures \ref{fignb} and \ref{fignc} display the profiles obtained when we assume a Lorentzian or Voigt-profile for $\Phi_{C'C}^{\mathrm{sp}}(\nu)$, respectively. The probability density function associated with a Voigt profile (i.e., a convolution of a Gaussian by a Lorentzian) is
\begin{equation}\label{voi}
    \mathscr{V}(\delta,\sigma,x)=\frac{e^{\frac{(\delta -i x)^2}{2 \sigma ^2}} \text{erfc}\left(\frac{\delta -i x}{\sqrt{2} \sigma }\right)+e^{\frac{(\delta +i x)^2}{2 \sigma ^2}} \text{erfc}\left(\frac{\delta +i x}{\sqrt{2} \sigma }\right)}{2 \sqrt{2 \pi } \sigma },
\end{equation}
where $\delta$ and $\sigma$ are the parameters of the Lorentzian and Gaussian functions respectively , and
\begin{equation}
    \mathrm{erfc}(x)=\frac{2}{\sqrt{\pi}}\int_{x}^{\infty}\mathrm{e}^{-t^{2}}\,\mathrm {d} t     
\end{equation}
is the complementary error function.

\section{Conclusion}\label{sec6}

We have investigated radiative rates between configurations, trying to remove, at least partially, some approximations (such as neglecting the energy spread) resulting from the usual averaging process. The formalism, inspired by Klapisch's work, \cite{Klapisch1993} is based on corrections involving the Unresolved-Transition-Array variance and shift. The corrections were computed for an iron plasma at $\rho$ = 0.17 g/cm$^3$ and $T$ = 182 eV, conditions similar to those at the boundary of the convective zone of the Sun. Under these conditions, the Rosseland mean opacity is primarily dominated by 2p-$n$d transitions with $n\geq 3, 4$. We have also studied an iron plasma at $\rho$ = 0.01 g/cm$^3$ and $T$ = 22 eV. Such conditions are typical of laser absorption-spectroscopy experiments that were carried out in recent decades and are roughly representative (at least in terms of temperature and mean ionization) of the conditions in the envelopes of $\beta-$Cephei type stars. Under these conditions, the dominant transitions are primarily $\Delta n=0$ ones, specifically within the $n=3$ shell. We have found that the corrections are more pronounced for 3-3 transitions than for 2p-$n$d ones. This is attributed to a greater overlap of wavefunctions in the first case, yielding a higher energy shift. Finally, we have discussed, in relation with the detailed balance, the Einstein coefficients and the average line profiles, an alternative approach for correcting the rates to guarantee the correct local-thermodynamic-equilibrium limit. 

We plan to study other transitions and elements such as oxygen, which plays also an important role in modeling the Sun and white dwarfs \cite{Mayes2025}, and rare earth elements \cite{Sarma2024}, for which atomic data are urgently needed by astronomers to study kilonovae emitted by neutron star mergers \cite{Kasen2017}. Additionally, we aim to study elements important for fusion research. As concerns magnetic-confinement fusion, tungsten is considered a leading candidate for divertor components in current tokamaks \cite{Hirai2007}, due to its high melting temperature, strong resistance to erosion, and limited capacity to trap tritium. However, if tungsten atoms enter the plasma, their densely populated emission lines can cause significant radiative energy losses \cite{Ralchenko2006}. Regarding inertial confinement fusion, germanium and silicon are used as dopants in the ablator \cite{Benredjem2013,Jarrah2017}. Following the present work, our first application will involve $n=4\rightarrow n=4$ transitions in tin for EUV lithography (see for instance Refs. \cite{AlRabban2006,Sheil2021,Sheil2023}). We also plan to investigate the case of super-transition arrays \cite{Hansen2011}, extending then the corrections to the superconfiguration formalism. This might be a difficult task, as it requires calculating, using canonical partition functions, averages over subshell populations of the present corrections, which contain terms that involve the inverse of powers of $E_{CC'}$. 



\bibliographystyle{unsrt}
\bibliography{biblio_klap93}

\appendix

\section{Numerical values of high-order moments up to \texorpdfstring{$\alpha_6$}{alpha6}}\label{appA}

The exact values of the moments up to $\alpha_6$ for 2p-3d and 2p-4d transition arrays for an iron plasma at $\rho=$0.17 g/cm$^3$ and $T=$182 eV are displayed in tables \ref{tab3} and \ref{tab5} respectively, and the same quantities for 3p-3d and 3s-3p transition arrays in an iron plasma at $\rho=$0.01 g/cm$^3$ and $T=$22 eV are given in tables \ref{tab8} and \ref{tab10}. The interpretation if $\alpha_5$ and $\alpha_6$ is more difficult than the one of $\alpha_3$ (``asymmetry of the distribution'') and $\alpha_4$ (``flattening/sharpness of the distribution''). The odd-order moments of the Gaussian are zero (since the distribution is symmetric), and the odd-order reduced centered moments are
\begin{equation*}
    \alpha_{2k}[\mathrm{Gaussian}]=\frac{(2k)!}{2^{k}k!}.
\end{equation*}

\begin{table}[!ht]
\begin{center}
\begin{tabular}{lrrrr}\hline\hline
Transition array & $\alpha_3$ & $\alpha_4$ & $\alpha_5$ & $\alpha_6$\\\hline\hline
2s$^2$ 2p$^5$ - 2s$^2$ 2p$^4$ 3d$^1$ & -0.4449 & 3.2427 & -2.9489 & 18.8368\\
2s$^2$ 2p$^4$ - 2s$^2$ 2p$^3$ 3d$^1$ & -0.57167 & 3.1849 & -5.0078 & 23.5106\\
2s$^2$ 2p$^3$ - 2s$^2$ 2p$^2$ 3d$^1$ & -0.7382 & 3.3843 & -6.4440 & 26.86145\\
2s$^1$ 2p$^5$ - 2s$^1$ 2p$^4$ 3d$^1$ & -0.2117 & 2.4176 & -0.7701 & 7.6216\\
2s$^2$ 2p$^6$ - 2s$^2$ 2p$^5$ 3d$^1$ & -1.5625 & 3.5849 & -7.9036 & 19.1082\\ 
2s$^1$ 2p$^4$ - 2s$^1$ 2p$^3$ 3d$^1$ & -0.3959 & 2.8813 & -2.7991 & 15.5853\\
2s$^1$ 2p$^3$ - 2s$^1$ 2p$^2$ 3d$^1$ & -0.5533 & 3.0480 & -4.5892 & 19.2346 \\
2s$^2$ 2p$^5$ 3d$^1$ - 2s$^2$ 2p$^4$ 3d$^2$ & -0.2019 & 3.0924 & -1.1466 & 17.7896\\
2s$^2$ 2p$^4$ 3d$^1$ - 2$s^2$ 2p$^3$ 3d$^2$ & -0.3281 & 3.0036 & -2.9695 & 19.7532\\
2s$^1$ 2p$^6$ - 2s$^1$ 2p$^5$ 3d$^1$ & 0.0007 & 1.2711 & -0.2858 & 1.9745\\
2s$^2$ 2p$^5$ 3p$^1$ - 2s$^2$ 2p$^4$ 3p$^1$ 3d$^1$ & -0.4030 & 3.8254 & -3.1620 & 27.0529\\
2s$^2$ 2p$^4$ 3p$^1$ - 2s$^2$ 2p$^3$ 3p$^1$ 3d$^1$ & -0.4882 & 3.6587 & -4.7620 & 28.40345\\
2s$^2$ 2p$^2$ - 2s$^2$ 2p$^1$ 3d$^1$ & -0.7681 & 3.4779 & -6.3476 & 22.3020\\
2s$^2$ 2p$^3$ 3d$^1$ - 2s$^2$ 2p$^2$ 3d$^2$ & -0.4605 & 3.1496 & -4.2238 & 22.1950\\
2s$^1$ 2p$^5$ 3d$^1$ - 2s$^1$ 2p$^4$ 3d$^2$ & -0.0733 & 2.3593 & 0.0126 & 7.6386\\
2s$^2$ 2p$^6$ 3d$^1$ - 2s$^2$ 2p$^5$ 3d$^2$ & -1.0064 & 2.8586 & -5.5880 & 14.0212\\
2s$^1$ 2p$^4$ 3d$^1$ - 2s$^1$ 2p$^3$ 3d$^2$ & -0.2513 & 2.7428 & -1.7007 & 13.8521\\
2s$^2$ 2p$^5$ 4f$^1$ - 2s$^2$ 2p$^4$ 4f$^1$ 3d$^1$ & -0.4627 & 3.3737 & -3.2588 & 20.1979\\
2s$^2$ 2p$^3$ 3p$^1$ - 2s$^2$ 2p$^2$ 3p$^1$ 3d$^1$ & -0.6081 & 3.82435 & -6.0709 & 31.6969\\
2s$^2$ 2p$^4$ 4f$^1$ - 2s$^2$ 2p$^3$ 4f$^1$ 3d$^1$ & -0.5492 & 3.2516 & -4.7724 & 22.6884\\\hline\hline
\end{tabular}
\caption{High-order moments of various 2p-3d transition arrays in an iron plasma at $\rho=$0.17 g/cm$^3$ and $T=$182 eV. The average skewness is $\alpha_3=-0.50391$ and the average kurtosis is $\alpha_4=3.08164$.}\label{tab3}
\end{center}
\end{table}

\begin{table}[!ht]
\begin{center}
\begin{tabular}{lrrrr}\hline\hline
Transition array & $\alpha_3$ & $\alpha_4$ & $\alpha_5$ & $\alpha_6$\\\hline\hline
2s$^2$ 2p$^5$ - 2s$^2$ 2p$^4$ 4d$^1$ & 0.5128 & 3.0875 & 4.5631 & 18.8366 \\
2s$^2$ 2p$^4$ - 2s$^2$ 2p$^3$ 4d$^1$ & 0.0974 & 2.3257 & 0.5777 & 10.1794\\
2s$^2$ 2p$^3$ - 2s$^2$ 2p$^2$ 4d$^1$ & -0.2869 & 2.4746 & -1.9854 & 11.4545\\
2s$^1$ 2p$^5$ - 2s$^1$ 2p$^4$ 4d$^1$ & 0.3292 & 2.1421 & 2.0118 & 6.7282\\
2s$^2$ 2p$^6$ - 2s$^2$ 2p$^5$ 4d$^1$ & -0.0649 & 1.0198 & -0.1584 & 1.0918\\
2s$^1$ 2p$^4$ - 2s$^1$ 2p$^3$ 4d$^1$ & 0.0271 & 2.2181 & 0.2400 & 8.4048 \\
2s$^1$ 2p$^3$ - 2s$^1$ 2p$^2$ 4d$^1$ & -0.2562 & 2.2904 & -1.6556 & 9.2872\\
2s$^2$ 2p$^5$ 3d$^1$ - 2s$^2$ 2p$^4$ 3d$^1$ 4d$^1$ & 0.4772 & 3.0847 & 4.1584 & 18.6603\\
2s$^2$ 2p$^4$ 3d$^1$ - 2s$^2$ 2p$^3$ 3d$^1$ 4d$^1$ & 0.1050 & 2.4351 & 0.5890 & 11.0356\\
2s$^1$ 2p$^6$ - 2s$^1$ 2p$^5$ 4d$^1$ & 0.6739 & 1.6893 & 1.8500 & 3.32545\\
2s$^2$ 2p$^5$ 3p$^1$ - 2s$^2$ 2$p^4$ 3$p^1$ 4d$^1$ & 0.5687 & 3.1108 & 4.4714 & 18.4952\\
2s$^2$ 2p$^4$ 3p$^1$ - 2s$^2$ 2p$^3$ 3p$^1$ 4d$^1$ & 0.1640 & 2.4154 & 0.8638 & 10.8656\\
2s$^2$ 2p$^2$ - 2s$^2$ 2p$^1$ 4d$^1$ & -0.5794 & 3.4577 & -6.0599 & 22.5402\\
2s$^2$ 2p$^3$ 3d$^1$ - 2s$^2$ 2p$^2$ 3d$^1$ 4d$^1$ & -0.2387 & 2.5495 & -1.8164 & 12.2014\\
2s$^1$ 2p$^5$ 3d$^1$ - 2s$^1$ 2p$^4$ 3d$^1$ 4d$^1$ & 0.3278 & 2.1818 & 1.9890 & 7.0782\\
2s$^2$ 2p$^6$ 3d$^1$ - 2s$^2$ 2p$^5$ 3d$^1$ 4d$^1$ & 0.0940 & 1.7723 & 0.6629 & 4.4407\\
2s$^1$ 2p$^4$ 3d$^1$ - 2s$^1$ 2p$^3$ 3d$^1$ 4d$^1$ & 0.0357 & 2.2550 & 0.1769 & 8.3298\\
2s$^2$ 2p$^5$ 4f$^1$ - 2s$^2$ 2p$^4$ 4f$^1$ 4d$^1$ & 0.4627 & 3.1057 & 4.2475 & 18.6458\\
2s$^2$ 2p$^3$ 3p$^1$ - 2s$^2$ 2p$^2$ 3p$^1$ 4d$^1$ & -0.1894 & 2.4853 & -1.4943 & 11.4970\\
2s$^2$ 2p$^4$ 4f$^1$ - 2s$^2$ 2p$^3$ 4f$^1$ 4d$^1$ & 0.0889 & 2.4049 & 0.5801 & 10.5482\\\hline\hline
\end{tabular}
\caption{High-order moments of various 2p-4d transition arrays in an iron plasma at $\rho=$0.17 g/cm$^3$ and $T=$182 eV. The average skewness is $\alpha_3=0.117445$ and the average kurtosis $\alpha_4=2.42529$. For the Gaussian assumption underlying the UTA formalism, one has $\alpha_3=\alpha_5=0$, $\alpha_4=3$ and $\alpha_6=15$.}\label{tab5}
\end{center}
\end{table}

\begin{table}[!ht]
\begin{center}
\begin{tabular}{lrrrr}\hline\hline
Transition array & $\alpha_3$ & $\alpha_4$ & $\alpha_5$ & $\alpha_6$\\\hline\hline
3s$^2$ 3p$^5$ 3d$^2$ - 3s$^2$ 3p$^4$ 3d$^3$ & -0.7570 & 5.1570 & -11.78335 & 65.5673\\
3s$^2$ 3p$^5$ 3d$^1$ - 3s$^2$ 3p$^4$ 3d$^2$ & -1.2016 & 6.7924 & -22.3603 & 119.0097\\
3s$^2$ 3p$^5$ 3d$^3$ - 3s$^2$ 3p$^4$ 3d$^4$ & -0.4975 & 5.0645 & -8.2631 & 58.97605\\
3s$^2$ 3p$^4$ 3d$^2$ - 3s$^2$ 3p$^3$ 3d$^3$ & -0.5995 & 5.3927 & -9.3716 & 59.2505\\
3s$^2$ 3p$^6$ 3d$^2$ - 3s$^2$ 3p$^5$ 3d$^3$ & -0.9638 & 3.9961 & -10.3136 & 35.79155\\
3s$^2$ 3p$^4$ 3d$^1$ - 3s$^2$ 3p$^3$ 3d$^2$ & -0.8514 & 5.9345 & -14.6166 & 79.7710\\
3s$^2$ 3p$^4$ 3d$^3$ - 3s$^2$ 3p$^3$ 3d$^4$ & -0.4415& 4.9889 & -6.2054 & 51.5826\\
3s$^2$ 3p$^6$ 3d$^1$ - 3s$^2$ 3p$^5$ 3d$^2$ & -1.5402 & 6.8054 & -28.4597 & 129.4351\\
3s$^2$ 3p$^6$ 3d$^3$ - 3s$^2$ 3p$^5$ 3d$^4$ & -0.6895 & 3.0658 & -5.6452 & 17.9990\\
3s$^2$ 3p$^5$ 3d$^4$ - 3s$^2$ 3p$^4$ 3d$^5$ & -0.3022 & 4.9557 & -6.1251 & 56.71855\\
3s$^2$ 3p$^5$ 3d$^0$ - 3s$^2$ 3p$^4$ 3d$^1$ & -2.9495 & 11.5221 & -46.67855 & 204.6961\\
3s$^2$ 3p$^4$ 3d$^4$ - 3s$^2$ 3p$^3$ 3d$^5$ & -0.1405& 5.1445 & -3.0569 & 57.4132\\
3s$^2$ 3p$^3$ 3d$^2$ - 3s$^2$ 3p$^2$ 3d$^3$ & -0.4070 & 4.8335 & -5.3561 & 44.0947\\ 
3s$^2$ 3p$^6$ 3d$^4$ - 3s$^2$ 3p$^5$ 3d$^5$ & -0.5081 & 2.5909 & -3.6092 & 11.5744\\
3s$^2$ 3p$^3$ 3d$^1$ - 3s$^2$ 3p$^2$ 3d$^2$ & -0.5427 & 5.4062 & -9.0423 & 57.2827\\
3s$^2$ 3p$^3$ 3d$^3$ - 3s$^2$ 3p$^2$ 3d$^4$ & -0.2979 & 4.7310 & -3.2112 & 43.2797\\ 
3s$^2$ 3p$^4$ 3d$^0$ - 3s$^2$ 3p$^3$ 3d$^1$ & -1.8211 & 6.3845 & -22.3393 & 88.5818\\ 
3s$^2$ 3p$^6$ 3d$^0$ - 3s$^2$ 3p$^5$ 3d$^1$ & -23.4431 & 558.0335 & -13504.57 & 333313.8\\
3s$^2$ 3p$^5$ 3d$^5$ - 3s$^2$ 3p$^4$ 3d$^6$ & -0.2201 & 4.2636 & -3.7001 & 40.2825\\ 
3s$^1$ 3p$^5$ 3d$^2$ - 3s$^1$ 3p$^4$ 3d$^3$ & -0.8968 & 5.2325 & -12.4910 & 61.2482\\\hline\hline
\end{tabular}
\caption{High-order moments of various 3p-3d transition arrays in an iron plasma at $\rho=$0.01 g/cm$^3$ and $T=$22 eV. The K shell is always full. The average skewness value is $\alpha_3=-1.95355$ and the average kurtosis value $\alpha_4=33.0148$. Excluding the pathologic very sharp array, one finds $\alpha_3=5.3822$. For the Gaussian assumption underlying the UTA formalism, one has $\alpha_3=\alpha_5=0$, $\alpha_4=3$ and $\alpha_6=15$.}\label{tab8}
\end{center}
\end{table}

\begin{table}[!ht]
\begin{center}
\begin{tabular}{lrrrr}\hline\hline
Transition array & $\alpha_3$ & $\alpha_4$ & $\alpha_5$ & $\alpha_6$\\\hline\hline
3s$^2$ 3p$^5$ 3d$^2$ - 3s$^1$ 3p$^6$ 3d$^2$ & -1.5117 & 5.2651 & -14.2063 & 45.13624\\
3s$^2$ 3p$^5$ 3d$^1$ - 3s$^1$ 3p$^6$ 3d$^1$ & -2.2632 & 8.7150 & -30.6312 & 111.8025\\
3s$^2$ 3p$^5$ 3d$^3$ - 3s$^1$ 3p$^6$ 3d$^3$ & -1.1179 & 4.0322 & -8.8090 & 26.7319\\
3s$^2$ 3p$^4$ 3d$^2$ - 3s$^1$ 3p$^5$ 3d$^2$ & -0.5153 & 3.9903 & -4.88495 & 28.1438\\ 
3s$^2$ 3p$^4$ 3d$^1$ - 3s$^1$ 3p$^5$ 3d$^1$ & -0.4489 & 4.1832 & -5.5541 & 30.95275\\
3s$^2$ 3p$^4$ 3d$^3$ - 3s$^1$ 3p$^5$ 3d$^3$ & -0.4050 & 3.6838 & -2.9445 & 24.0619\\
3s$^2$ 3p$^5$ 3d$^4$ - 3s$^1$ 3p$^6$ 3d$^4$ & -0.8459 & 3.3910 & -5.9148 & 18.33365\\
3s$^2$ 3p$^5$ 3d$^0$ - 3s$^1$ 3p$^6$ 3d$^0$ & -0.7071 & 1.5000 & -1.7678 & 2.7500\\
3s$^2$ 3p$^4$ 3d$^4$ - 3s$^1$ 3p$^5$ 3d$^4$ & -0.3243 & 3.5368 & -2.8136 & 23.2579\\
3s$^2$ 3p$^3$ 3d$^2$ - 3s$^1$ 3p$^4$ 3d$^2$ & -0.3511 & 3.9708 & -1.9859 & 28.4896\\
3s$^2$ 3p$^3$ 3d$^1$ - 3s$^1$ 3p$^4$ 3d$^1$ & -0.3892 & 3.5567 & -3.7447 & 23.4798\\
3s$^2$ 3p$^3$ 3d$^3$ - 3s$^1$ 3p$^4$ 3d$^3$ & -0.3128 & 3.8585 & -1.8968 & 28.0137\\
3s$^2$ 3p$^4$ 3d$^0$ - 3s$^1$ 3p$^5$ 3d$^0$ & 0.5172 & 1.7054 & 1.4109 & 4.5027\\
3s$^2$ 3p$^5$ 3d$^5$ - 3s$^1$ 3p$^6$ 3d$^5$ & -0.6277 & 2.9899 & -3.9440 & 13.5920\\
3s$^1$ 3p$^5$ 3d$^2$ - 3s$^0$ 3p$^6$ 3d$^2$ & -0.1763 & 3.3018 & -0.3012 & 16.6144\\\hline\hline
\end{tabular}
\caption{High-order moments of various 3s-3p transition arrays in an iron plasma at $\rho=$0.01 g/cm$^3$ and $T=$22 eV. The average skewness is equal to $\alpha_3=-0.631947$ and the average kurtosis to $\alpha_4=3.84537$. For the Gaussian assumption underlying the UTA formalism, one has $\alpha_3=\alpha_5=0$, $\alpha_4=3$ and $\alpha_6=15$.}\label{tab10}
\end{center}
\end{table}

\end{document}